\documentclass[english]{revtex4}
\usepackage[T1]{fontenc}
\usepackage[latin9]{inputenc}
\usepackage{amsthm}
\usepackage{amsmath}
\usepackage{amssymb}
\usepackage{graphicx}

\makeatletter
\@ifundefined{textcolor}{}
{%
 \definecolor{BLACK}{gray}{0}
 \definecolor{WHITE}{gray}{1}
 \definecolor{RED}{rgb}{1,0,0}
 \definecolor{GREEN}{rgb}{0,1,0}
 \definecolor{BLUE}{rgb}{0,0,1}
 \definecolor{CYAN}{cmyk}{1,0,0,0}
 \definecolor{MAGENTA}{cmyk}{0,1,0,0}
 \definecolor{YELLOW}{cmyk}{0,0,1,0}
}

\usepackage{esint}

\makeatletter
\@ifundefined{textcolor}{}
{%
 \definecolor{BLACK}{gray}{0}
 \definecolor{WHITE}{gray}{1}
 \definecolor{RED}{rgb}{1,0,0}
 \definecolor{GREEN}{rgb}{0,1,0}
 \definecolor{BLUE}{rgb}{0,0,1}
 \definecolor{CYAN}{cmyk}{1,0,0,0}
 \definecolor{MAGENTA}{cmyk}{0,1,0,0}
 \definecolor{YELLOW}{cmyk}{0,0,1,0}
 }

\usepackage{subfig}
\usepackage{graphicx}
\usepackage{color}

\makeatother

\makeatother

\usepackage{babel}
\begin{document}

\title{Kinetic microtearing modes and reconnecting modes in strongly magnetised
slab plasmas}

\author{A Zocco$^{1,2,3}$, N F Loureiro$^{4}$, D Dickinson$^{1,5},$ R Numata$^{6}$,
C M Roach$^{1}$ }

\address{$^{1}$Culham Centre for Fusion Energy, Culham Science Centre, Abingdon,
Oxon, OX14 3DB, UK}

\address{$^{2}$ Rudolf Peierls Centre for Theoretical Physics, 1 Keble Road,
Oxford, OX1 3NP, UK}

\address{$^{3}$Max-Planck-Institut für Plasmaphysik, Wendelsteinstrasse,
D-17489, Greifswald, Germany}

\address{$^{4}$Instituto de Plasmas e Fusão Nuclear, Instituto Superior T\'ecnico,
Universidade de Lisboa, 1049-001, Lisboa, Portugal}

\address{$^{5}$York Plasma Institute, Dept.
of Physics, University of York, Heslington York YO10 5DD, UK}
\address{$^{6}$Graduate School of Simulation Studies, University of Hyogo
7-1-28 Minatojima Minami-machi, Chuo-ku, Kobe, Hyogo, 650-0047}

\begin{abstract}
The problem of the linear microtearing mode in a slab magnetised plasma,
and its connection to kinetic reconnecting modes, is addressed. Electrons
are described using a novel hybrid fluid-kinetic model that captures
electron heating, ions are gyrokinetic. Magnetic reconnection can
occur as a result of either electron conductivity and inertia, depending
on which one predominates. We eschew the use of an energy dependent
collision frequency in the collisional operator model, unlike previous
works. A model of the electron conductivity that matches the weakly
collisional regime to the exact Landau result at zero collisionality
and gives the correct electron isothermal response far from the reconnection
region is presented. We identify in the breaking of the constant-$A_{\parallel}$
approximation the necessary condition for microtearing instability
in the collisional regime. Connections with the theory of collisional
non-isothermal (or semicollisional) and collisionless tearing-parity
electron temperature gradient driven (ETG) modes are elucidated. 
\end{abstract}
\maketitle

\section{Introduction\label{sec:Introduction}}

The presence of microtearing modes in fusion plasmas was predicted
by Hazeltine, Dobrott and Wang \citep{hazeltine:1778} in 1975$^{1}$\footnotetext[1]{For
an historical introduction see Ref. \citep{0741-3335-32-10-004}.}.
These modes are driven unstable by the electron temperature gradient,
and rotate in the electron direction with a real frequency of the
order of the electron drift frequency. They can drive magnetic reconnection
even for magnetic equilibria that are tearing-mode stable.

Microtearing modes have been found unstable in most magnetic confinement
systems \citep{stallard:1978,PhysRevLett.92.235003,0029-5515-39-2-303,0741-3335-49-8-001,0741-3335-53-3-035013,PhysRevLett.106.155004,guttenfelder:022506,PhysRevLett.105.195001,0741-3335-52-4-045007,0741-3335-51-12-124020}.
In some tokamak experiments \citep{0029-5515-51-7-073045}, nonlinear
microtearing physics seems to be the key to explaining the favourable
scaling of energy confinement with collisionality. There is also growing
evidence \citep{PhysRevLett.106.155004,PhysRevLett.106.155003,doerk:055907,PhysRevLett.108.235002,PhysRevLett.108.135002,PhysRevLett.110.155005}
that the simple electrostatic picture of tokamak gyrokinetic turbulence
could be inappropriate in many regimes that are operationally relevant,
and microtearing certainly plays a role in this. Microtearing activity
has also been detected in a reversed-field-pinch configuration \citep{zuin}. 

In gyrokinetic simulations, microtearing modes almost inevitably manifest
themselves when electromagnetic effects are considered. Their phenomenology
is rather complicated, and has been summarised in Ref. \citep{0741-3335-55-7-074006}.
Amongst recent authors involved in microtearing research, some are
re-proposing the idea \citep{PhysRevLett.44.994} that the mode is
responsible for anomalous electron transport in magnetic fusion devices,
which is believed to be predominantly electromagnetic \citep{PhysRevLett.106.155003,PhysRevLett.108.235002,doerk:055907}.
To understand better the physics of the mode in conditions relevant
to a fusion reactor, microtearing studies have been extended to include
toroidal effects \citep{PhysRevLett.108.135002,0741-3335-55-7-074006,1.4799980},
finite $\beta$ (the ratio of plasma kinetic to magnetic pressure)
\citep{guttenfelder:022506,0741-3335-55-7-074006,doerk:055907}, realistic
mass ratios \citep{guttenfelder:022506}, and more or less sophisticated
model collision operators \citep{0741-3335-49-8-001,PhysRevLett.106.155003}.
In the literature, we can also find claims of quantitative agreement
between predicted and measured levels of transport (see Ref. \citep{PhysRevLett.99.135003}
for example). Surprisingly, despite this renaissance in the study
of the mode, a slab theory which retains electrostatic perturbations,
and small but finite collisionality is still missing. In this work
we present such a theory. We avoid imposing constant magnetic perturbations
across the reconnection region, a simplification known as ``constant-$A_{\parallel}$
approximation'' \citep{FKR} used in most analytical previous works
\citep{hazeltine:1778,drake:1341,chang-dominguez-hazeltine,Rosenberg,Gladdmicrotearing,10.1063/1.863352,10.1063/1.863576,0741-3335-32-10-004}.
Full ion Larmor orbit effects \citep{antonsen-coppi} are retained,
even for non-constant magnetic perturbations \citep{0741-3335-28-4-003,pegoraro:364,0741-3335-54-3-035003}.
Finite $\hat{\beta}_{T}=0.5\beta_{e}L_{s}^{2}/L_{T}^{2}$ effects
are also retained \citep{drake:2509,0741-3335-52-4-045007}. Here
$\beta_{e}$ is the ratio of electron to magnetic pressure, whereas
$L_{s}$ and $L_{T}$ are the characteristic magnetic shear and electron
temperature gradient scales, respectively. Indeed, as we will prove,
a finite $\hat{\beta}_{T}$ theory is required to describe unstable
microtearing modes. Analytical progress can be made if we combine
the approaches introduced in Ref. \citep{0741-3335-54-3-035003} and
in Ref. \citep{zocco:102309} for ions and electrons, respectively.
The first approach is based on the separation between electron and
ion scales, and crucially on some results borrowed from the theory
of generalized functions. The second is based on a spectral representation
of the electron distribution function in Hermite series \citep{zocco:102309},
that leads to a solution of the electron kinetic equation expressed
in terms of a continued fraction, reminiscent of earlier theories
\citep{Gladdmicrotearing,Rosenberg}. We do not make use of an energy
dependent collision frequency in the collision operator model. This
is crucial to generating a time-dependent thermal force in the parallel
momentum equation of the electrons that produces the instability in
the highly collisional limit, according to previous studies \citep{Gladdmicrotearing}.
However, it is unnecessary in our kinetic theory. In fact, we introduce
a new closure for the electron kinetic problem that allows us to study
the low and high collisionality limits. We show how our analytical
solution of the electron kinetic problem connects to the exact analytical
result at zero collisionality obtained using Landau contour integration.
The new electron solution reproduces the correct isothermal electron
response far from the reconnecting region. We find a dispersion relation
that relates the microtearing mode to the collisional, non-isothermal
(semicollisional), drift-tearing mode at high collisionality. We identify
in the breaking of the constant-$A_{\parallel}$ approximation \citep{FKR,migl}
the necessary condition for instability in the collisional regime.
We also prove that the coupling of the drift-tearing mode branch to
the kinetic Alfv\'en wave provides a mechanism to avoid the cancellation
of the microtearing mode drive that would happen when the collision
frequency is energy independent. This new mechanism that drives unstable
the microtearing mode is not excluding other mechanisms already present
in the literature such as the energy dependence of the collision frequency.

In the weakly collisional limit, we carry out a numerical analysis
with the gyrokinetic code {\tt GS2}. The only electron temperature
gradient driven weakly collisional reconnecting mode that we find
unstable is the tearing-parity, strongly driven ETG, which happens
to be mostly electrostatic. In the semicollisional regime, on the
other hand, we identify the microtearing mode and give the explicit
analytic expression for its growth rate and real frequency in the
cases of large and finite electron temperature gradients.

The paper is organised as follows: In Section II we present the nonlinear
and linear model equations. In Section II C, we solve the electron
kinetic equation, introduce the new electron conductivity and show
how it relates to previous theories in the semicollisional and truly
collisionless$(\nu_{ei}\equiv0)$ limits. In Section III we derive
a new dispersion relation for drift-kinetic reconnecting modes with
gyrokinetic ions. In Section IV and V we study the collisional and
weakly collisional limits of drift-kinetic reconnecting modes, and
benchmark our results against hybrid fluid-kinetic and gyrokinetic
codes. In Section VI we report on our new results on the theory of
kinetic microtearing modes. In Section VII a numerical investigation
of the existence of collisionless microtearing modes is carried out.
Conclusions are presented in Section VIII.

\section{Model equations\label{sec:Model-equations}}

\subsection{Nonlinear model}

For our analysis, a generalisation of the hybrid fluid-kinetic model
derived in Ref. \citep{zocco:102309} is used. Electron temperature
gradients are introduced with a maximal ordering that allows us to
neglect the electron \textsl{and} ion drift frequencies $\omega_{*}\propto L_{n_{0}}^{-1},$
where $L_{n_{0}}$ is the characteristic background density gradient
length scale. We consider 
\begin{equation}
\omega\sim\nu_{ei}\sim k_{\parallel}v_{the}\sim k_{\parallel}v_{A}\sim\omega_{T},\label{eq:ordering}
\end{equation}
where $\omega_{T}=\eta_{e}\omega_{*}\equiv0.5k_{y}v_{the}\rho_{e}/L_{T},$
and $L_{T}^{-1}=T_{0e}^{-1}\partial_{x}T_{0e}$ defines the characteristic
temperature gradient length scale, $v_{the}$ the electron thermal
speed, $v_{A}$ the Alfv\'en speed, $\omega$ the mode frequency,
$k_{\parallel}$ the wave number parallel to the magnetic field, and
$\nu_{ei}$ the electron-ion collision frequency. In this way we have
\begin{equation}
\omega\frac{h_{e}}{F_{0e}}\sim k_{y}\rho_{e}\frac{v_{the}}{L_{T_{e}}}\frac{e\varphi}{T_{0e}}\sim\frac{v_{the}}{L_{T_{e}}}\varepsilon,\label{eq:ordering2}
\end{equation}
 where $\rho_{e}$ is the electron Larmor radius, $k_{y}$ the mode
wave number perpendicular to the magnetic field, and $\varepsilon=k_{\parallel}/k_{\perp}\sim h_{e}/F_{0e}$
is the small expansion parameter of gyrokinetic theory, with $k_{\perp}^{2}=k_{y}^{2}+k^{2},$
where $k$ is the radial wave vector, and $h_{e}$ the non-adiabatic
part of the electron distribution function. The ordering in Eq. (\ref{eq:ordering2})
is derived from the original ordering of Ref. \citep{zocco:102309},
where the electrostatic potential, $\varphi,$ is ordered as $e\varphi/T_{0e}\sim\varepsilon/\sqrt{\beta_{e}},$
with $\beta_{e}=8\pi n_{0e}T_{0e}/B_{0}^{2}$, and $\beta_{e}\sim(m_{e}/m_{i}).$
The ordering in $\beta_{e}$ is necessary to include electron inertia
and accommodate the kinetic Alfv\'en wave \citep{zocco:102309}.
From Eq. (\ref{eq:ordering}), it follows that $k_{\parallel}L_{T}\sim1,$
therefore we set $L_{n_{0e}}/L_{T}\equiv\eta_{e}\sim\beta_{e}^{-1/2}\sim(m_{i}/m_{e})^{1/2}\gg1.$
This is simply a flat density limit, with $\omega_{*}\rightarrow0$
but $\omega_{T}=\eta_{e}\omega_{*}\equiv\mathcal{O}(1).$ Since electrons
are drift-kinetic, $k_{\perp}\rho_{e}\sim\sqrt{\beta_{e}}\sim\sqrt{m_{e}/m_{i}}\ll1,$
the electron diamagnetic drift frequency can only be important, compared
to the Alfv\'en frequency, $\omega_{A}=v_{A}/L_{s}\equiv\tau_{A}^{-1}$,
at very low shear; in fact 
\[
\frac{\omega_{T}}{\omega_{A}}=\sqrt{\frac{\beta_{e}}{2}}\frac{L_{s}}{L_{T}}k_{y}\rho_{s},
\]
 where $L_{s}$ is the magnetic shear length, $\rho_{s}=\sqrt{Z/(2\tau)}\rho_{i},$
$\tau=T_{0i}/T_{0e},$ and $Z$ is the charge number. The parameter
\[
\hat{\beta}_{T}\equiv(\beta_{e}/2)L_{s}^{2}/L_{T}^{2}
\]
 is the familiar parameter of semicollisional theory \citep{drake:2509,0741-3335-54-3-035003}.
In the following, for ease of comparison of our results with previous
results, we will sometimes leave explicit the combination $\rho_{e}/L_{T};$
however that should always be regarded as $\rho_{e}/L_{T}\equiv d_{e}/L_{s}\sqrt{\beta_{e}}L_{s}/L_{T},$
with $\sqrt{\beta_{e}}L_{s}/L_{T}\sim\mathcal{O}(1),$ where $d_{e}=c/\omega_{pe}\propto\sqrt{m_{e}}$
is the electron inertial length with $\omega_{pe}=(4\pi n_{0e}e^{2}/m_{e})^{1/2}$
the electron plasma frequency.

Using these orderings, the equations that we obtain can be listed
below; their detailed derivation can be found in Appendix A. They
are: the electron continuity equation, 
\begin{equation}
\frac{d}{dt}\frac{Z}{\tau}\left(1-\hat{\Gamma}_{0}\right)\frac{e\varphi}{T_{0e}}=\hat{\mathbf{b}}\cdot\nabla\frac{e}{m_{e}c}d_{e}^{2}\nabla_{\perp}^{2}A_{\parallel},\label{eq:elcont}
\end{equation}
 with $\Gamma_{0}=I_{0}(k^{2}\rho_{i}^{2}/2)\exp[-k^{2}\rho_{i}^{2}/2],$
where $I_{0}$ is the modified Bessel function (the ``hat'' is symbolic
for the inverse Fourier transform), $\hat{\mathbf{b}}\cdot\nabla=-\partial_{z}+B_{0}^{\text{-1}}\{A_{\parallel},\},$
$\{,\}$ is the Poisson bracket, and $A_{\text{\ensuremath{\parallel}}}$
the parallel component of the magnetic potential, the generalised
Ohm's law, 
\begin{equation}
\begin{split} & \frac{d}{dt}(A_{\parallel}-d_{e}^{2}\nabla_{\perp}^{2}A_{\parallel})=-c\frac{\partial\varphi}{\partial z}+\frac{T_{0e}c}{e}\hat{\mathbf{b}}\cdot\nabla\left[\frac{Z}{\tau}\left(\hat{\Gamma}_{0}-1\right)\frac{e\varphi}{T_{0e}}+\frac{\delta T_{\parallel e}}{T_{0e}}\right]\\
 & +\eta\nabla_{\perp}^{2}A_{\parallel}-\frac{1}{2}\frac{\rho_{e}}{L_{T}}v_{the}\frac{\partial}{\partial y}A_{\parallel},
\end{split}
\label{eq:grnohmslaw}
\end{equation}
 with 

\begin{equation}
\frac{\delta T_{\parallel e}}{T_{0e}}\equiv\frac{1}{n_{0e}}\int d^{3}\mathbf{v}2\frac{v_{\parallel}^{2}}{v_{the}^{2}}g_{e},\label{eq:delT}
\end{equation}
and the kinetic equation 
\begin{equation}
\begin{split} & \frac{dg_{e}}{dt}+v_{\parallel}\left[\hat{\mathbf{b}}\cdot\nabla g_{e}-F_{0e}\hat{\mathbf{b}}\cdot\nabla\frac{\delta T_{\parallel e}}{T_{0e}}\right]-C[g_{e}]=\\
 & F_{0e}\left(1-2\frac{v_{\parallel}^{2}}{v_{the}^{2}}\right)\hat{\mathbf{b}}\cdot\nabla\frac{e}{m_{e}c}d_{e}^{2}\nabla_{\perp}^{2}A_{\parallel}+\\
 & -\frac{1}{2}\frac{\rho_{e}}{L_{T}}v_{the}\frac{\partial}{\partial y}\left[\left(\frac{m_{e}v_{\parallel}^{2}}{2T_{0e}}-\frac{1}{2}\right)\frac{e\varphi}{T_{0e}}-2\frac{v_{\parallel}}{v_{the}}\left(\frac{m_{e}v_{\parallel}^{2}}{2T_{0e}}-\frac{3}{2}\right)\frac{e}{cm_{e}}\frac{A_{\parallel}}{v_{the}}\right]F_{0e}.
\end{split}
\label{eq:sum3}
\end{equation}
 Here $g_{e}$ is a kinetic function such that the non-adiabatic part
of the electron distribution function can be written as 
\begin{equation}
h_{e}=\left(-\frac{e\varphi}{T_{0e}}+\frac{\delta n_{e}}{n_{0e}}+\frac{v_{\parallel}u_{\parallel e}}{T_{0e}}m_{e}\right)F_{0e}+g_{e}+\mathcal{O}\left(\frac{m_{e}}{m_{i}}\right),\label{eq:split}
\end{equation}
 with $\int dv_{\parallel}(1,v_{\parallel})g_{e}=0.$ The collision
operator is defined as 
\begin{equation}
C[g_{e}]=\left(\frac{\partial h_{e}}{\partial t}\right)_{coll}-2\frac{v_{\parallel}F_{0e}}{v_{the}^{2}n_{oe}}\int d^{3}\mathbf{v}v_{\parallel}\left(\frac{\partial h_{e}}{\partial t}\right)_{coll}.
\end{equation}
A collision operator model will be introduced when necessary. Equation
(\ref{eq:sum3}), in the limit of homogeneous backgrounds, reduces
to the result of Ref. \citep{zocco:102309}. All undefined symbols
are standard.

\subsection{Linear eigenvalue problem \label{sub:Linear-eigenvalue-problem}}

Consider the following magnetic configuration 
\begin{equation}
\mathbf{B}=B_{0}\hat{\mathbf{z}}+\delta B_{y}^{(0)}(x)\hat{\mathbf{y}}+\delta\mathbf{B}_{\perp}^{(1)},\label{eq:magneq}
\end{equation}
 where $\delta B_{\perp}^{(1)}\ll\delta B_{y}^{(0)}\ll B_{0},$ and
$\delta B_{y}^{(0)}(x)$ is part of the perturbation of the guide
field $B_{0}.$ Then $\delta B_{y}^{(0)}=-dA_{\parallel}^{(0)}/dx\equiv B_{0}f(x).$The
function $f$ does not need to be specified at this stage. Let us
consider for a moment the solution of Eq. (\ref{eq:sum3}) that will
be crucial in our following analysis {[}see Eqs. (\ref{eq:tempeq})
and (\ref{eq:timedep_ETC}) later{]}. We linearize Eqs. (\ref{eq:elcont})-(\ref{eq:grnohmslaw})-(\ref{eq:sum3})
around (\ref{eq:magneq}) to obtain 
\begin{equation}
\omega\frac{\delta n_{e}}{n_{0e}}=k_{\parallel}(x)u_{\parallel e},\label{eq:lin1}
\end{equation}

\begin{equation}
\begin{split} & -i\frac{\omega}{c}\left[1-\frac{\omega_{T}}{\omega}\right]A_{\parallel}+ik_{\parallel}(x)\varphi=\\
 & ik_{\parallel}(x)\frac{T_{0e}}{e}\left(\frac{\delta n_{e}}{n_{0e}}+\frac{\delta T_{\parallel e}}{T_{0e}}\right)+\frac{\eta-i\omega d_{e}^{2}}{c}\left(\partial_{x}^{2}-k_{y}^{2}\right)A_{\parallel},
\end{split}
\label{eq:lin2}
\end{equation}
 and 
\begin{equation}
\begin{split} & \frac{\delta T_{\parallel e}}{T_{0e}}=\frac{2k_{\parallel}(x)u_{\parallel e}}{\omega+i\kappa_{\parallel e}(\omega,\nu_{ei})k_{\parallel}^{2}(x)}\\
 & +\frac{\kappa_{\parallel e}(\omega)k_{\parallel}(x)}{\omega+i\kappa_{\parallel e}(\omega,\nu_{ei})k_{\parallel}^{2}(x)}i\frac{\omega_{T}}{c}\frac{eA_{\parallel}}{T_{0e}}+\frac{\omega_{T}}{\omega+i\kappa_{\parallel e}(\omega,\nu_{ei})k_{\parallel}^{2}(x)}\frac{e\varphi}{T_{0e}},
\end{split}
\label{eq:lin3}
\end{equation}
 where $\kappa_{\parallel e}=\kappa_{\parallel e}(\omega,\nu_{ei})$
is a frequency dependent parallel electron thermal conductivity, $k_{\parallel}(x)=k_{y}f(x),$
and we have neglected $d_{e}^{2}f^{\prime\prime}(x)/f(x)$ compared
to $d_{e}^{2}\partial_{x}^{2},$ and assumed $k_{z}\ll k_{y}f(x).$
After using Eqs. (\ref{eq:lin2}) and (\ref{eq:lin3}) in Eq. (\ref{eq:lin1})
we obtain 
\begin{equation}
u_{\parallel e}=-\frac{e}{m_{e}\left(\nu_{ei}-i\omega\right)}\left[i\frac{\omega}{c}A_{\parallel}-ik_{\parallel}(x)\varphi\right]\sigma_{e},\label{eq:parelsc}
\end{equation}
 where $\sigma_{e}$ is the electron conductivity. The explicit forms
of $\sigma_{e}$ and $\kappa_{\parallel e}$ will be given shortly.
If we use $f(x)\approx x/L_{s}$ in the neighborhood of the surface
at which $k_{\parallel}(x)=0,$ (the equilibrium magnetic field is
sheared with characteristic length $L_{s}$), then we obtain the following
set of equations for kinetic reconnecting drift modes: 
\begin{equation}
-\frac{x}{\delta}\left(A_{\parallel}-\frac{x}{\delta}\hat{\varphi}\right)\sigma_{e}(x/\delta)=\left(1-i\frac{\omega}{\nu_{ei}}\right)\int_{-\infty}^{\infty}dpe^{ipx}F(p\rho_{i})\hat{\varphi}(p),\label{eq:semi1}
\end{equation}

\begin{equation}
\frac{1}{\hat{\omega}^{2}\hat{\beta}}\left(\frac{d^{2}}{dx^{2}}-k_{y}^{2}\right)A_{\parallel}=\frac{1}{\delta^{2}}\frac{\delta}{x}\int_{-\infty}^{\infty}dpe^{ipx}F(p\rho_{i})\hat{\varphi}(p),\label{eq:semi2}
\end{equation}
 where $\hat{\varphi}=(\delta/L_{s})(k_{y}c/\omega)\varphi,$ $\hat{\omega}=\omega/\omega_{T},$
$F=(Z/\tau)(\Gamma_{0}-1)$ , and
\begin{equation}
\delta^{2}=\exp[-i\pi/2]2\omega\nu_{e}/(k_{y}^{2}v_{the}^{2})L_{s}^{2}
\end{equation}
is the electron semicollisional scale.

Equations (\ref{eq:semi1}) and (\ref{eq:semi2}) can be solved using
a double asymptotic matching technique. By using the separation between
ion and electron scales, $\delta\ll\rho_{i},$ the two equations can
be simplified in the two regions where $x\sim\delta\ll L_{s},$ and
$x\sim\rho_{i}\ll L_{s},$ and matched in the overlapping regions
to give a dispersion relation for the mode. It turns out that, when
this separation of scales is allowed for, it is possible to formulate
the problem as a single eigenvalue equation for the current density
$J=-\partial_{x}^{2}A_{\parallel}$ \citep{0741-3335-28-4-003}. A
technical difficulty arises since the electron region equation is
naturally formulated (and solved) in real space, whereas for the ion
region, $x\sim\rho_{i},$ a formulation in Fourier space is more convenient.
Nevertheless, the problem of identifying and matching the correct
solutions was solved in Ref. \citep{0741-3335-54-3-035003}, and the
same approach will be fruitful here. The case $k_{z}\gg k_{y}f(x)$
will be analysed in the truly collisionless limit, $\nu_{ei}\equiv0,$
after integrating exactly Eq. (\ref{eq:sum3}).

\subsection{Electron conductivity\label{sub:Electron-conductivity}}

In Ref. \citep{zocco:102309}, a Hermite expansion of the electron
distribution function $g_{e}$ was introduced. This allowed us to
prove the Boltzmann H-theorem, and predict the velocity space spectrum
of electron free energy in steady-state and in the presence of growing
modes. The use of Hermite polynomials as a basis in velocity space
also proved useful for the numerical implementation of the new kinetic
description of the electrons (also known as Kinetic Reduced Electron
Heating Model) \citep{zocco:102309,loureiro-scecco-zocco}. Here,
we find that such a representation is again very powerful. In this
section, we give details of the new equations for inhomogeneous backgrounds,
and show their implications in the collisional and weakly collisional
limit. In particular, we calculate explicitly the electron conductivity
$\sigma_{e}$ introduced in Eqs. (\ref{eq:semi1}) and (\ref{eq:semi2}).

\subsubsection{Hermite Series}

If we aim at describing the parallel velocity space dynamics of our
kinetic system, we are allowed to use a simple model collision operator.
As in Ref. \citep{zocco:102309}, it is useful to employ the (modified)
Lenard-Bernstein collision operator\citep{lenard:1456,zocco:102309}
\begin{equation}
C[g_{e}]=\nu_{ei}\left[\frac{1}{2}\frac{\partial}{\partial\hat{v}_{\parallel}}\left(\frac{\partial}{\partial\hat{v}_{\parallel}}+2\hat{v}_{\parallel}\right)g_{e}-(1-2\hat{v}_{\parallel}^{2})\frac{\delta T_{\parallel e}}{T_{0e}}F_{0e}\right],\label{eq:LBcoll}
\end{equation}
where $\nu_{ei}=4\pi n_{0e}Z^{2}e^{4}\ln\Lambda/(m_{e}^{2}v_{the}^{3})$
is the energy-independent collision frequency. Let us project Eq.
(\ref{eq:sum3}) onto Hermite polynomials$^{1}$ \footnotetext[1]{Since
the model has no knowledge of perpendicular temperature fluctuations,
we consider them fixed to an arbitrary constant. As a consequence,
we modify the operator to conserve parallel temperature fluctuations.
This is not consistent for $\nu_{ei}\sim\omega,$ and $\omega\sim\omega_{T},$
however it is correct in the subsidiary limits considered here $\nu_{ei}\gg\omega,$
and $N^{-1/2}\ll\nu_{ei}/\omega\ll1,$ where $N$ is the highest Hermite
moment kept. Other authors have recognised the role of conservation
of electron parallel temperature perturbations during collisions in
order to obtain a microtearing instability {[}see App. A of Ref. \citep{Rosenberg}{]}.},
which are eigenfunctions of the collision operator model in Eq. (\ref{eq:LBcoll}).
The resulting equation is 
\begin{equation}
\begin{split} & \frac{1}{n_{0e}}\frac{d}{dt}n_{0e}\hat{g}_{m}+\frac{1}{n_{0e}}v_{the}\hat{\mathbf{b}}\cdot\nabla n_{0e}\left(\sqrt{\frac{m+1}{2}}\hat{g}_{m+1}+\sqrt{\frac{m}{2}}\hat{g}_{m-1}-\delta_{m,1}\hat{g}_{2}\right)\\
 & =-\sqrt{2}\delta_{m,2}\left(\hat{\mathbf{b}}\cdot\nabla u_{\parallel e}+\frac{\eta_{e}}{2}\frac{\mathbf{v}_{E}\cdot\nabla n_{0e}}{n_{0e}}\right)\\
 & -\sqrt{3}\delta_{m,3}\frac{\eta_{e}}{2}\frac{v_{the}\tilde{\mathbf{b}}\cdot\nabla n_{0e}}{n_{0e}}-\nu_{ei}(m\hat{g}_{m}-2\delta_{m,2}),
\end{split}
\label{eq:exp}
\end{equation}
 where $\tilde{\mathbf{b}}\equiv-B_{0}^{-1}\{A_{\parallel},\cdots\}.$
Here $\hat{g}_{e}=2v_{the}^{-2}\int dv_{\perp}v_{\perp}\exp[-v_{\perp}^{2}/v_{the}^{2}]g_{e}$,
whereas the Hermite inverse transform is defined as 
\begin{equation}
\hat{g}_{e}(v_{\parallel})=\sum_{m=0}^{\infty}\frac{H_{m}(\hat{v}_{\parallel})}{\sqrt{2^{m}m!}}\hat{g}_{m}F_{0e}(\hat{v}_{\parallel}^{2}),
\end{equation}
 with coefficients 
\begin{equation}
\hat{g}_{m}=\frac{1}{n_{0e}}\int_{-\infty}^{\infty}d\hat{v}_{\parallel}\frac{H_{m}(\hat{v}_{\parallel})}{\sqrt{2^{m}m!}}\hat{g}_{e}(v_{\parallel}),
\end{equation}
 where $\hat{v}_{\parallel}=v_{\parallel}/v_{the}.$ Hence, for the
first Hermite moments we obtain 
\begin{equation}
\begin{split} & \frac{d}{dt}\hat{g}_{2}+v_{the}\frac{\hat{\mathbf{b}}\cdot\nabla}{n_{0e}}\left(\sqrt{\frac{3}{2}}n_{0e}\hat{g}_{3}\right)\\
 & =-\sqrt{2}\left(\hat{\mathbf{b}}\cdot\nabla u_{\parallel e}+\frac{\eta_{e}}{2}\frac{\mathbf{v}_{E}\cdot\nabla n_{0e}}{n_{0e}}\right),
\end{split}
\label{eq:g2equation}
\end{equation}
 for $m=2,$ 
\begin{equation}
\begin{split} & \frac{d}{dt}\hat{g}_{3}+v_{the}\frac{\hat{\mathbf{b}}\cdot\nabla}{n_{0e}}\left(\sqrt{2}n_{0e}\hat{g}_{4}+\sqrt{\frac{3}{2}}n_{0e}\hat{g}_{2}\right)=\\
 & -\sqrt{3}\frac{\eta_{e}}{2}\frac{v_{the}\tilde{\mathbf{b}}\cdot\nabla n_{0e}}{n_{0e}}-3\nu_{ei}\hat{g}_{3},
\end{split}
\label{eq:g3}
\end{equation}
 for $m=3$ and 
\begin{equation}
\frac{d}{dt}\hat{g}_{m}+\frac{1}{n_{0e}}v_{the}\hat{\mathbf{b}}\cdot\nabla n_{0e}\left(\sqrt{\frac{m+1}{2}}\hat{g}_{m+1}+\sqrt{\frac{m}{2}}\hat{g}_{m-1}\right)=-m\nu_{ei}\hat{g}_{m},\label{eq:mmaggiroe4}
\end{equation}
 for $m\geq4.$

\subsubsection{Collisional limit $\nu_{ei}\gg\omega$}

In this limit the Hermite coefficients scale as \citep{zocco:102309}
\begin{equation}
\frac{\hat{g}_{m}}{\hat{g}_{m-1}}\sim\frac{k_{\parallel}v_{the}}{\sqrt{m}\nu_{ei}},
\end{equation}
and we can truncate the fluid system by neglecting $\hat{g}_{4}$
in the $\hat{g}_{3}$ equation and invert $\hat{g}_{3}$ from Eq.
(\ref{eq:g3}), neglecting the time derivative compared to the collision
frequency. In this way, we obtain 
\begin{equation}
\hat{g}_{3}\approx-\frac{1}{3\nu_{ei}}\frac{v_{the}}{n_{0e}}\hat{\mathbf{b}}\cdot\nabla\sqrt{\frac{3}{2}}n_{0e}\hat{g}_{2}-\frac{\sqrt{3}}{3\nu_{ei}}\frac{\eta_{e}}{2}\frac{1}{n_{0e}}v_{the}\tilde{\mathbf{b}}\cdot\nabla n_{0e}.\label{eq:collclosure}
\end{equation}
 The resulting equation for the temperature perturbation is 
\begin{equation}
\begin{split} & \frac{1}{n_{0e}}\frac{d}{dt}n_{0e}\frac{\delta T_{\parallel e}}{T_{0e}}=\frac{v_{the}}{n_{0e}}\hat{\mathbf{b}}\cdot\nabla\frac{v_{the}}{2\nu_{ei}}\hat{\mathbf{b}}\cdot\nabla n_{0e}\frac{\delta T_{\parallel e}}{T_{0e}}\\
 & +\frac{\eta_{e}}{2}\frac{v_{the}}{n_{0e}}\hat{\mathbf{b}}\cdot\nabla\frac{v_{the}}{\nu_{ei}}\frac{\tilde{\mathbf{b}}\cdot\nabla n_{0e}}{n_{0e}}\\
 & -\eta_{e}\frac{\mathbf{v}_{E}\cdot\nabla n_{0e}}{n_{0e}}-2\hat{\mathbf{b}}\cdot\nabla u_{\parallel e}.
\end{split}
\label{eq:tempeq}
\end{equation}
 Equation (\ref{eq:tempeq}) is coupled to Ohm's law (\ref{eq:grnohmslaw})
via $\delta T_{\parallel e}$ in the collisional limit. In this case,
electron inertia can be neglected compared to the collisional term,
we can use Eq. (\ref{eq:tempeq}) in Ohm's law and obtain the conductivity
for collisional, non-isothermal electrons

\begin{equation}
\sigma_{e}(x/\delta)=\frac{\sigma_{0}+\frac{x^{2}}{\delta^{2}}}{1+4\frac{x^{2}}{\delta^{2}}+\frac{x^{4}}{\delta^{4}}},\label{eq:cond}
\end{equation}
 with 
\begin{equation}
\sigma_{0}=1-\frac{\omega_{T}}{\omega}.
\end{equation}
The numerical coefficients in Eq. (\ref{eq:cond}) differ from those
of fluid theories \citep{drake-phfl-84-rmhd} because of the details
of the collisional operator model. This proves mathematically that
our model reproduces the resuts of Ref. \citep{0741-3335-54-3-035003}
in the highly collisional limit.

\subsubsection{General electron conductivity}

An alternative way of closing the kinetic hierarchy is by considering

\begin{equation}
\frac{\hat{g}_{m}}{\hat{g}_{m-1}}\sim\frac{k_{\parallel}v_{the}}{\sqrt{m}\omega}\frac{\omega}{\nu_{ei}}\ll1,\label{eq:closMT-1}
\end{equation}
 with 
\begin{equation}
\frac{k_{\parallel}v_{the}}{\omega\sqrt{m}}\ll1,\,\,\,\mbox{but}\,\,\,\frac{k_{\parallel}v_{the}}{\omega}\sim\frac{\omega}{\nu_{ei}}\sim1,
\end{equation}
thus the expansion is in large $m\gg1.$ This closure scheme works
also nonlinearly. Our velocity space representation allows us to avoid
taking the $\omega/\nu_{ei}\ll1$ limit in order to calculate velocity
space integrals, and allows us to study the interesting and realistic
limit $N^{-1/2}\ll\nu_{ei}/\omega\ll1$, where $N$ is the order of
the highest Hermite moment kept. 

Let us consider an $N\gg1$ for which $\hat{g}_{N+1}\ll\hat{g}_{N}$
in the sense of Eq. (\ref{eq:closMT-1}). Indeed, there is always
one for small and finite $\nu_{ei}.$ Then, for the $Nth$ component
the kinetic equation is 
\begin{equation}
(-i\omega+N\nu_{ei})\hat{g}_{N}=-ik_{\parallel}v_{the}\sqrt{\frac{N}{2}}\hat{g}_{N-1}.\label{eq:closure}
\end{equation}
 We can use this expression for $\hat{g}_{N}$ in the equation for
the $N-1$ component and obtain the $N-1$ component as a function
of the $N-2$ component 
\begin{equation}
\left[-i\omega+\left(N-1\right)\nu_{ei}+ik_{\parallel}v_{the}\frac{-ik_{\parallel}v_{the}N/2}{-i\omega+N\nu_{ei}}\right]\hat{g}_{N-1}=-ik_{\parallel}v_{the}\sqrt{\frac{N-1}{2}}\hat{g}_{N-2}.
\end{equation}
 Hence, after $n$ iterations we have 
\begin{equation}
\begin{split} & \hat{g}_{N-n}=-ik_{\parallel}v_{the}\sqrt{\frac{N-n}{2}}\hat{g}_{N-(n+1)}\times\\
 & \frac{1}{\left[-i\omega+\left(N-n\right)\nu_{ei}\right]+ik_{\parallel}v_{the}\frac{-ik_{\parallel}v_{the}(N-n+1)/2}{\left[-i\omega+\left(N-n+1\right)\nu_{ei}\right]+ik_{\parallel}v_{the}\frac{-ik_{\parallel}v_{the}(N-n+2)/2}{\cdots+ik_{\parallel}v_{the}\frac{-ik_{\parallel}v_{the}N/2}{-i\omega+N\nu_{ei}}}}}
\end{split}
.\label{eq:gensolkin}
\end{equation}
 Now, when $N-n=4,$ we are able to write $\hat{g}_{3}$ in Eq. (\ref{eq:g2equation})
explicitly as a function of all other $\hat{g}_{m}$ up to $\hat{g}_{N}.$
By proceeding in the same way as in the collisional case, we obtain
a general electron conductivity, 
\begin{equation}
\hat{\sigma}_{e}=\frac{\sigma_{0}+\frac{3\nu_{ei}}{-i\omega+3\nu_{ei}+\frac{4}{2}\frac{k_{\parallel}^{2}v_{the}^{2}}{\Omega(N)}}s^{2}}{1+\left[\frac{3}{1-i\frac{\omega}{\nu_{ei}}}+\frac{3\nu_{ei}}{-i\omega+3\nu_{ei}+\frac{4}{2}\frac{k_{\parallel}^{2}v_{the}^{2}}{\Omega(N)}}\right]s^{2}+\frac{1}{1-i\frac{\omega}{\nu_{ei}}}\frac{3\nu_{ei}}{-i\omega+3\nu_{ei}+\frac{4}{2}\frac{k_{\parallel}^{2}v_{the}^{2}}{\Omega(N)}}s^{4}},\label{eq:conduttivita}
\end{equation}
 where $s^{2}=k_{\parallel}^{2}\kappa_{\parallel e}^{f}/(-i\omega)\equiv x^{2}/\delta^{2},$
$\kappa_{\parallel e}^{f}=v_{the}^{2}/(2\nu_{ei})$ is the fluid parallel
electron thermal conductivity, and 
\begin{equation}
\Omega(N)=\left[-i\omega+4\nu_{ei}\right]+ik_{\parallel}v_{the}\frac{-ik_{\parallel}v_{the}5/2}{\left[-i\omega+5\nu_{ei}\right]+ik_{\parallel}v_{the}\frac{-ik_{\parallel}v_{the}6/2}{\cdots+ik_{\parallel}v_{the}\frac{-ik_{\parallel}v_{the}N/2}{-i\omega+N\nu_{ei}}}}.
\end{equation}
In the limit $\omega/\nu_{ei}\ll1$ Eq. (\ref{eq:conduttivita}) reduces
to the semicollisional electron conductivity (\ref{eq:cond}). Equation
(\ref{eq:conduttivita}) implies that the electron thermal conductivity
is the product of the fluid part times a kinetic contribution that
encapsulates the time evolution of all the Hermite moments kept, that
is

\begin{equation}
\kappa_{\parallel e}(\omega)=\kappa_{\parallel e}^{f}\frac{3\nu_{ei}}{-i\omega+3\nu_{ei}+ik_{\parallel}v_{the}\frac{-ik_{\parallel}v_{the}4/2}{\cdots+ik_{\parallel}v_{the}\frac{-ik_{\parallel}v_{the}N/2}{-i\omega+N\nu_{ei}}}}.\label{eq:timedep_ETC}
\end{equation}

If we compare Eq. (\ref{eq:cond}) with Eq. (13) of Ref. \citep{drake:2509},
we see that the term proportional to the microtearing drive, $\omega_{T}/\omega,$
is a different $\mathcal{O}(1)$ number. This is due to the fact that
the model collisional operator, as such, can give correct results
up to $\mathcal{O}(1)$ multiplicative constants. In previous theories
of microtearing modes, the frequency dependence of \textit{this} term,
originated by a time dependent electron thermal force, has been proposed
as crucial to obtain an instability. That is, an electron conductivity
of the form
\begin{equation}
\sigma_{D}=\frac{1-\frac{\omega_{T}}{\omega}\left(1+\alpha+i\alpha\alpha^{\prime}\frac{\omega}{\nu_{ei}}\right)+d_{1}s^{2}}{1+d_{0}s^{2}+d_{1}s^{4}}\label{eq:modelcond}
\end{equation}
was used. Here $\omega\ll\nu_{ei},$ $d_{0},$ $d_{1},$ $\alpha,$
and $\alpha^{\prime}$ are positive real constants. In this formulation,
the term proportional to $\alpha^{\prime}$ is responsible for the
microtearing instability. If we take the same $\omega\ll\nu_{ei}$
limit of $(1-i\omega/\nu_{ei})^{-1}\hat{\sigma}_{e},$ which enters
Eqs. (\ref{eq:semi1}) and (\ref{eq:semi2}), and retain the small
collisional correction that couples to $\omega_{T},$ we obtain 
\begin{equation}
(1-i\omega/\nu_{ei})^{-1}\hat{\sigma}_{e}\approx\frac{1-\frac{\omega_{T}}{\omega}\left(1+i\frac{\omega}{\nu_{ei}}\right)+s^{2}}{1+4s^{2}+s^{4}}.\label{eq:concond}
\end{equation}
Then, the electron inertial term $(1-i\omega/\nu_{ei})^{-1}$ is coupled
to the electron temperature gradient as the time dependent electron
thermal force contribution of previous works. Notice that the approximation
for the electron conductivity originally used by Hazeltine et al.
\citep{hazeltine:1778}, would give an electron temperature gradient
instability in the weakly collisional limit $\nu_{ei}/\omega\ll1$
\citep{Rosenberg}, as pointed out by Rosenberg and co-authors \citep{Rosenberg},
while we expect the mode to be marginally stable in the collisionless
limit. In any case, the conductivity, calculated in Ref. \citep{10.1063/1.863576}
to obtain the generalised spatial dependence of the conductivity originally
used by Hazeltine et al. \citep{hazeltine:1778} {[}Eq. (27){]}, does
not give the correct isothermal electron response at large distances
from the reconnecting layer, that is, $s^{2}(1-i\omega/\nu_{ei})^{-1}\hat{\sigma}_{e}\neq1,$
for $s\gg1$ \citep{pegoraro:364,PhysRevLett.66.425,0741-3335-54-3-035003}.
Therefore, it is not appropriate to properly match the electron solution
to the ion region solution in a theory with large ion Larmor orbits
\citep{pegoraro:364,PhysRevLett.66.425,0741-3335-54-3-035003}.

As already noticed by Drake et al. \citep{drake:2509}, in the theory
of semicollisional drift tearing modes, the use of a model for the
electron conductivity {[}as in Eq. (\ref{eq:modelcond}){]}, instead
of the fluid one of Eq. (\ref{eq:cond}), brings about an imaginary
correction to the fundamental frequency of the mode which gives an
instability in leading order.

While we leave open the question whether the use of such a model can
predict results from first principle numerical simulations, we content
ourselves with the simple model collisional operator in Eq. (\ref{eq:LBcoll}),
and aim at deriving a theory that simultaneously takes account of
electron inertia, non-isothermal electrons, the electrostatic potential,
and non-constant magnetic perturbations.

\subsubsection{Finite collisionality and collisionless limit}

\begin{figure}[!h]
\subfloat{(a)}{\includegraphics[scale=0.8]{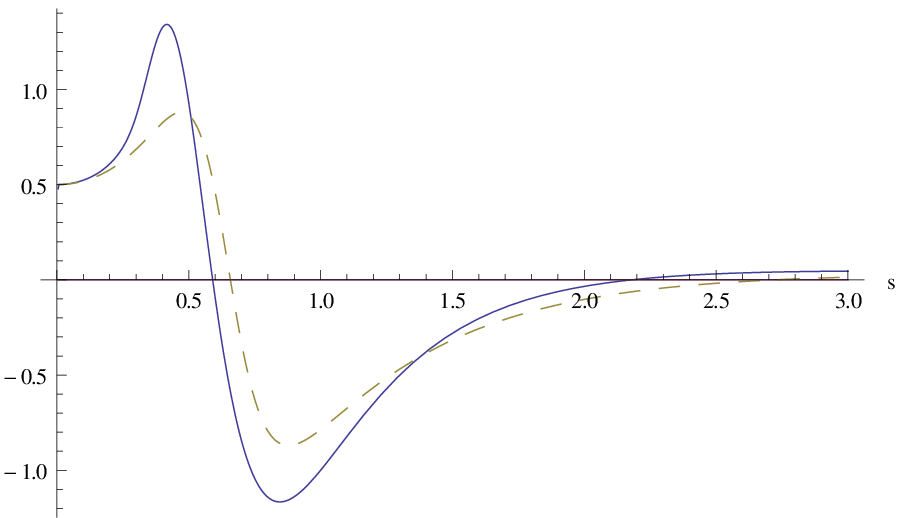}} \subfloat(b){\includegraphics[scale=0.8]{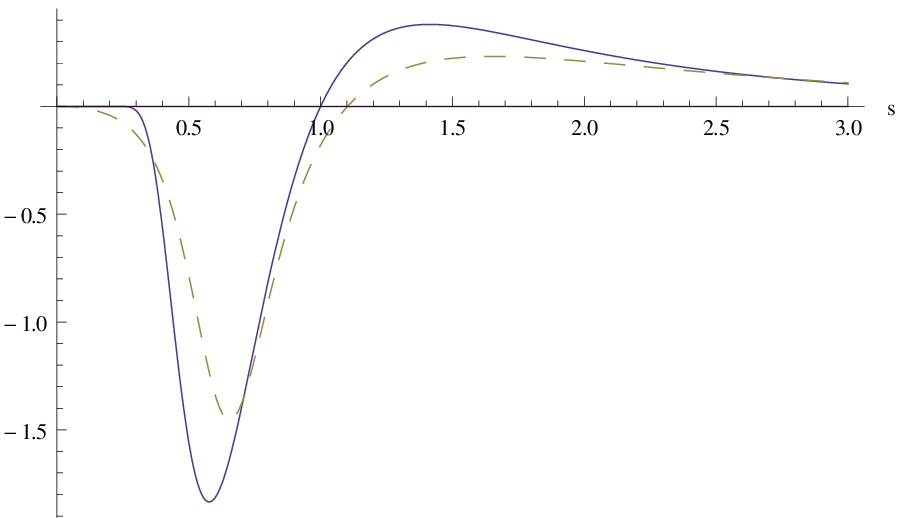}} 

\caption{The real (a) and imaginary (b) parts of the electron conductivity
$\hat{\sigma}_{L}$ in the $\nu_{ei}=0$ case (solid line) calculated
using Eq. (\ref{eq:land cond}), and $-1/2\hat{\sigma}_{e}$ for $N=100$
(dashed line) calculated using Eq. (\ref{eq:conduttivita}). Here
$\omega/\nu_{ei}=0.5/0.1$ and $s\equiv x/\delta_{c-less}.$ }

\label{fig:sigmaM6} 
\end{figure}

\begin{figure}[!h]
\subfloat{(a)}{\includegraphics[scale=0.8]{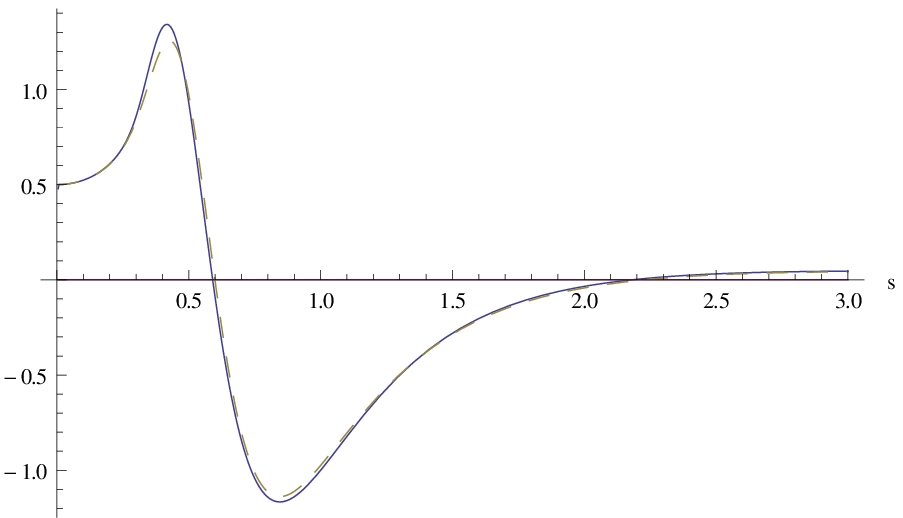}} \subfloat{(b)}{\includegraphics[scale=0.8]{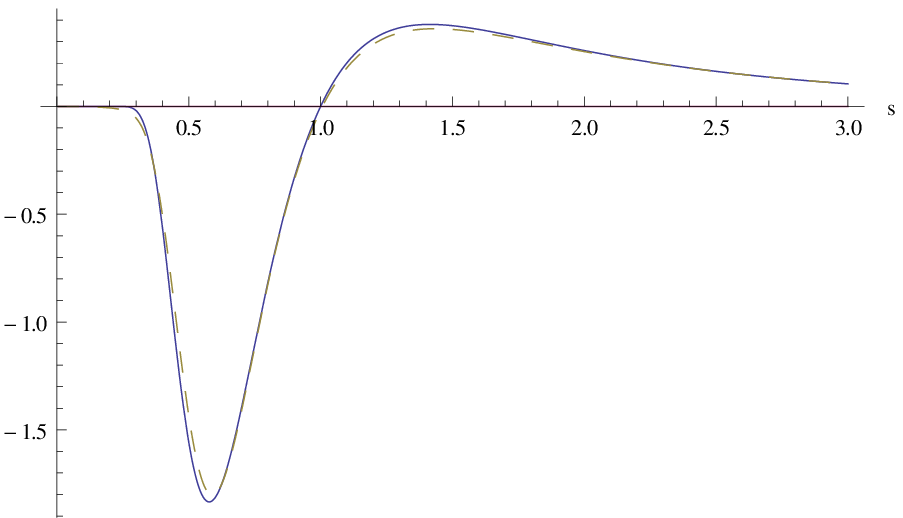}} 

\caption{The real (a) and imaginary (b) parts of the electron conductivity
$\hat{\sigma}_{L}$ in the $\nu_{ei}=0$ case (solid line) calculated
using Eq. (\ref{eq:land cond}), and $-1/2\hat{\sigma}_{e}$ for $N=100$
(dashed line) calculated using Eq. (\ref{eq:conduttivita}). Here
$\omega/\nu_{ei}=0.5/0.01$ and $s\equiv x/\delta_{c-less}.$}

\label{fig:sigmaM20} 
\end{figure}

If we were to neglect collisions completely, we could solve the electron
kinetic equation Eq. (\ref{eq:sum3}) using Landau integrals \citep{landau}
and obtain \citep{coppi-collisionless,kadomtsev-pogutse,zocco:102309}
\begin{equation}
\hat{\sigma}_{L}\left(\frac{x}{\delta_{c-less}}\right)=-\frac{1}{2}\left\{ \frac{\delta_{c-less}^{2}}{x^{2}}Z^{\prime}\left(\frac{\delta_{c-less}}{|x|}\right)+\frac{1}{2\hat{\omega}}\frac{\delta_{c-less}^{3}}{|x|^{3}}Z^{\prime\prime}\left(\frac{\delta_{c-less}}{|x|}\right)\right\} ,\label{eq:land cond}
\end{equation}
where $\delta_{c-less}=(\omega/k_{y}v_{the})L_{s},$ $s=x/\delta_{c-less},$
and $Z$ is the plasma dispersion function \citep{plasma-disp}. We
can compare the analytical form of the electron conductivity calculated
using Landau integrals, and the one calculated using the continued
fraction solution (\ref{eq:conduttivita}) for $N=100,$ $\nu_{ei}/\omega=0.1/0.5,$
and $\nu_{ei}/\omega=0.01/0.5.$ For this value of $N,$ the continued
fraction solution has converged in both cases. The agreement improves
with decreasing collisionality, see Figs. (\ref{fig:sigmaM6}) and
(\ref{fig:sigmaM20}). 
In the following, we will make use of the finite-collisionality formulation,
where all scales are normalised to the semicollisional scale $\delta^{2}=\exp[-i\pi/2]2\omega\nu_{ei}/(k_{y}^{2}v_{the}^{2})L_{s}^{2},$
{[}Eq. (16){]} and the electron conductivity is given by Eq. (\ref{eq:conduttivita}).
This turns out to be extremely convenient for numerical and analytical
purposes.

\section{Low$-\hat{\beta}_{T}$ and low$-k_{y}\delta$ Dispersion Relation\label{sec:Dispersion-Relation}}

To derive the low$-\hat{\beta}_{T},$ low$-k_{y}\delta$ dispersion
relation, we follow Ref. \citep{0741-3335-54-3-035003}. A general
dispersion relation can be written as 
\begin{equation}
\frac{\hat{c}_{-}}{\hat{c}_{+}}=\frac{\hat{a}_{-}}{\hat{a}_{+}}\frac{\delta}{\rho_{i}},\label{eq:gendisprel}
\end{equation}
 where the coefficients $\hat{a}_{\pm}$ are such that the large asymptotic
limit of the solution for the current $J$ is 
\begin{equation}
J(k\rho_{i})\sim\hat{a}_{+}k\rho_{i}+\hat{a}_{-},\,\,\,\mbox{for}\, k\rho_{i}\gg1,
\end{equation}
 in the ion region $x\sim\rho_{i}$, and 
\begin{equation}
J(k\delta)\sim\hat{c}_{+}k\delta+\hat{c}_{-},\,\,\,\mbox{for}\, k\delta\ll1,
\end{equation}
 in the electron region $x\sim\delta.$

In the electron region when $x\sim\delta\ll\rho_{i},$ we can use
the $k_{\perp}\rho_{i}\gg1$ limit for the RHS of Eqs. (\ref{eq:semi1})
and (\ref{eq:semi2}), to obtain \citep{0741-3335-54-3-035003} 
\begin{equation}
\frac{d^{2}}{ds^{2}}\left[\frac{\left(1-i\frac{\omega}{\nu_{ei}}\right)F_{\infty}-s^{2}\hat{\sigma}_{e}(s)}{F_{\infty}\hat{\sigma}_{e}(s)}\right]J(s)=-\hat{\beta}_{T}\hat{\omega}^{2}J(s),\label{eq:curr}
\end{equation}
 where $\hat{\omega}=\omega/\omega_{T},$ $s=x/\delta,$ and $F_{\infty}=-Z/\tau.$
Equation (\ref{eq:curr}) is valid for any collision frequency provided
$N^{-1/2}\ll\nu_{ei}/\omega\ll1,$ or $\nu_{ei}\gg\omega,$ with $N$
the order of the highest Hermite moment kept in the model. We retain
electron inertia in Ohm's law. For this reason we have a new term
$\left(1-i\omega/\nu_{ei}\right)F_{\infty}-s^{2}\hat{\sigma}_{e}(s),$
and not $F_{\infty}-s^{2}\hat{\sigma}_{e}(s),$ which was used in
the collisional case of Ref. \citep{0741-3335-54-3-035003}.

In order to derive the dispersion relation, we need to study the large
argument asymptotic behaviour of the solution of Eq. (\ref{eq:curr}).
This is determined by 
\begin{equation}
\frac{d^{2}}{ds^{2}}s^{2}J\sim-\hat{\beta}_{T}\hat{\omega}^{2}\frac{F_{\infty}}{F_{\infty}-1}J.\label{eq:whatever}
\end{equation}
 Equation (\ref{eq:whatever}) tells us that the solution of Eq. (\ref{eq:curr})
behaves asymptotically as 
\begin{equation}
J(s)\sim b_{+}s^{-1}+b_{-}s^{-2},\,\,\,\mbox{for}\, s\rightarrow\infty,\label{eq:ascurr}
\end{equation}
 so in $t-$space (the Fourier conjugate of $s$) we shall have 
\begin{equation}
J(t)\sim\hat{c}_{+}t^{1}+\hat{c}_{-}t^{0},\,\,\,\mbox{for}\, t\rightarrow0,
\end{equation}
 with \citep{gelfand,0741-3335-54-3-035003} 
\begin{equation}
\frac{\hat{c}_{-}}{\hat{c}_{+}}=\frac{\Gamma(\mu-\frac{1}{2})}{\Gamma(-\mu-\frac{1}{2})}\tan\left[\frac{\pi}{2}\left(\frac{1}{2}+\mu\right)\right]\frac{b_{+}}{b_{-}},\label{eq:formula}
\end{equation}
 and 
\begin{equation}
\frac{1}{4}-\mu^{2}=\hat{\omega}^{2}\hat{\beta}_{T}\frac{F_{\infty}}{G_{\infty}},
\end{equation}
 where $G_{\infty}=F_{\infty}-1.$ This result is general and \textsl{does
not} depend on the electron collision model used. Indeed, far from
the reconnection region, electrons are isothermal \citep{pegoraro:364,PhysRevLett.66.425,0741-3335-54-3-035003,zocco:102309}.
When we solve Eq. (\ref{eq:curr}) in  a low$-\hat{\beta}_{T}$ expansion,
we notice that the power $s^{-1}$ in Eq. (\ref{eq:ascurr}) is not
coming from the zeroth order solution of Eq. (\ref{eq:curr}). It
is therefore sufficient to solve it to first order. Indeed, we obtain
the reconnecting (even) solution 
\begin{equation}
J(s)=\frac{F_{\infty}\hat{\sigma}_{e}(s)}{\left(1-i\frac{\omega}{\nu_{ei}}\right)F_{\infty}-s^{2}\hat{\sigma}_{e}(s)}\left\{ 1-\hat{\beta}_{T}\hat{\omega}^{2}\int_{0}^{s}ds^{\prime}\int_{0}^{s^{\prime}}du\frac{F_{\infty}\hat{\sigma}_{e}(u)}{\left(1-i\frac{\omega}{\nu_{ei}}\right)F_{\infty}-u^{2}\hat{\sigma}_{e}(u)}\right\} ,\label{eq:firstorder}
\end{equation}
 and the large argument asymptotic behaviour is 
\begin{equation}
J(s)\sim\frac{1}{s^{2}}+\hat{\beta}_{T}\hat{\omega}^{2}I_{e}\frac{1}{s},
\end{equation}
 with 
\begin{equation}
\begin{split}I_{e} & =-\int_{0}^{\infty}ds\frac{F_{\infty}\hat{\sigma}_{e}(s)}{\left(1-i\frac{\omega}{\nu_{ei}}\right)F_{\infty}-s^{2}\hat{\sigma}_{e}(s)}.\end{split}
\label{eq:Ie}
\end{equation}
 Since the matching to the ion solution is performed in Fourier space,
in principle we should calculate the Fourier transform of Eq. (\ref{eq:firstorder}).
However, we can apply the analytical formula in Eq. (\ref{eq:formula})
\citep{0741-3335-54-3-035003} that relates asymptotic leading order
coefficients in real space with those in $k-$space \citep{gelfand}.
Then, in the small $\hat{\beta}_{T}$ limit (equivalently $\mu\rightarrow1/2^{+}$),
we have 
\begin{equation}
\frac{\hat{c}_{-}}{\hat{c}_{+}}\approx-\frac{2}{\pi}\frac{G_{\infty}}{F_{\infty}}I_{e}.
\end{equation}
 The ion region is treated in the same way as in Ref. \citep{0741-3335-54-3-035003}.
In the limit $x\gg\delta,$ the product $s^{2}\hat{\sigma}_{e}$ tends
to a constant, and we obtain a differential equation for the current
in Fourier space \citep{0741-3335-54-3-035003}. This can again be
solved in a low$-\hat{\beta}_{T}$ expansion as in Ref. \citep{0741-3335-54-3-035003}.
We report here the result 
\begin{equation}
\begin{split} & J(k)\sim k^{\hat{\beta}_{T}\hat{\omega}^{2}\frac{F_{\infty}}{G_{\infty}}}+\hat{\beta}_{T}\hat{\omega}^{2}\frac{F_{\infty}}{G_{\infty}}\frac{\pi}{\Delta^{\prime}\rho_{i}}\left\{ k+\frac{1}{G_{\infty}}\frac{1}{\sqrt{\pi}}\log k\right\} \\
 & +\hat{\beta}_{T}\hat{\omega}^{2}\frac{\pi}{\Delta^{\prime}\rho_{i}}\bar{I}-\left(\hat{\beta}_{T}\hat{\omega}^{2}\right)^{2}\frac{F_{\infty}}{G_{\infty}}k\int_{0}^{\infty}dk\frac{F}{k^{2}G},
\end{split}
\label{eq:ionregsol}
\end{equation}
 where $k\equiv k\rho_{i},$ and $\bar{I}=\int_{0}^{\infty}dk\left\{ F/G-F_{\infty}/G_{\infty}-(F_{\infty}/G_{\infty}^{2})\pi^{-1/2}/(1+k)\right\} ,$
and 
\[
\Delta^{\prime}=\left.\frac{1}{A_{\parallel}^{ext}}\frac{dA_{\parallel}^{ext}}{dx}\right|_{0^{-}}^{0^{+}}
\]
is the ideal MHD external solution parameter. So, the coefficients
$\hat{a}_{\pm}$ in Eq. (\ref{eq:gendisprel}) are 
\begin{equation}
\frac{\hat{a}_{-}}{\hat{a}_{+}}=\frac{1+\hat{\beta}_{T}\hat{\omega}^{2}\frac{F_{\infty}}{G_{\infty}^{2}}\frac{\pi}{\Delta^{\prime}\rho_{i}}\frac{1}{\sqrt{\pi}}\log\frac{\rho_{i}}{\delta}+\hat{\beta}_{T}\hat{\omega}^{2}\frac{\pi}{\Delta^{\prime}\rho_{i}}\bar{I}}{\hat{\beta}_{T}\hat{\omega}^{2}\frac{F_{\infty}}{G_{\infty}}\frac{\pi}{\Delta^{\prime}\rho_{i}}-\left(\hat{\beta}_{T}\hat{\omega}^{2}\right)^{2}\frac{F_{\infty}}{G_{\infty}}\int_{0}^{\infty}\frac{F}{k^{2}G}}
\end{equation}
From this, it follows that the dispersion relation is

\begin{equation}
\frac{\delta}{\rho_{i}}B(\hat{\omega})-\frac{2}{\pi}I_{e}\hat{\omega}^{2}C(\hat{\omega})=0,\label{eq:disprelfinal}
\end{equation}
 with 
\begin{equation}
B(\hat{\omega})=\frac{\Delta^{\prime}\rho_{i}}{\pi\hat{\beta}_{T}}-\hat{\omega}^{2}\frac{Z/\tau}{\left[Z/\tau+1\right]^{2}}\frac{1}{\sqrt{\pi}}\log\frac{\rho_{i}}{\delta}+\hat{\omega}^{2}\bar{I}(\tau),
\end{equation}
 and 
\begin{equation}
C(\hat{\omega})=1-\frac{\hat{\beta}_{T}\Delta^{\prime}\rho_{i}}{\pi}\hat{\omega}^{2}I(\tau).\label{Peg}
\end{equation}
 The ion integrals are defined as 
\begin{equation}
\bar{I}(\tau)=\int_{0}^{\infty}dq\left[\frac{F}{G}-\frac{Z/\tau}{Z/\tau+1}+\frac{Z/\tau}{\left[Z/\tau+1\right]^{2}}\frac{1}{\sqrt{\pi}}\frac{1}{(1+q)}\right],
\end{equation}
 and 
\begin{equation}
I(\tau)=\int_{0}^{\infty}\frac{dq}{q^{2}}\frac{F}{G}.
\end{equation}

We notice that the ion region solution, Eq. (\ref{eq:ionregsol}),
has already been matched to a boundary condition at $k\rho_{i}\rightarrow0,$
\citep{0741-3335-28-4-003} and the information about the ideal MHD
external solution is embedded in the parameter $\Delta^{\prime}.$
The authors of Ref. \citep{0741-3335-28-4-003} proved analytically
that the ideal MHD boundary condition used in (\ref{eq:ionregsol})
is correct even for high wave numbers. This generally corresponds
to the substitution $\Delta^{\prime}\rightarrow-2k_{y},$ when the
external solution is of the form $A_{\parallel}^{ext}\sim e^{-k_{y}\left|x\right|},$
and is tearing mode stable. Equation (\ref{eq:disprelfinal}), for
positive $\Delta^{\prime},$ is valid even for $\Delta^{\prime}\rho_{i}\gg1,$
that is when the so-called constant-$A_{\parallel}$ approximation
does not hold. In the present work, we will consider two situations:
arbitrary values of $\left|\Delta^{\prime}\rho_{i}\right|$ for $\Delta^{\prime}\rho_{i}>0,$
and $\Delta^{\prime}\rho_{i}<0,$ with $\left|\Delta^{\prime}\rho_{i}\right|\ll1.$
A thorough investigation of non-constant tearing-stable perturbations
is left to future work . For $\Delta^{\prime}\rho_{i}<0,$ with $\left|\Delta^{\prime}\rho_{i}\right|\ll1,$
we will also find it useful to match directly the electron region
solution to an exponentially decaying external solution.

In what follows, we first study Eq. (\ref{eq:disprelfinal}) in the
highly collisional limit $\nu_{ei}/\omega\gg1$; we then benchmark
our results against numerical hybrid and kinetic simulations and we
finally investigate microtearing modes and ETG modes within our theoretical
framework.

\section{Collisional limit\label{sec:Collisional-limit}}

\subsection{Collisional non-isothermal electrons}

By deriving Eq. (\ref{eq:cond}), we proved that, in the collisional
limit, the model considered here {[}Eqs. (\ref{eq:elcont})-(\ref{eq:grnohmslaw})-(\ref{eq:sum3}){]}
reproduces the results of Ref. \citep{0741-3335-54-3-035003}. We
then take $\omega_{*}\rightarrow0,$ but $\omega_{*}\eta_{e}\equiv\omega_{T}\sim\mathcal{O}(1),$
in Eq. (33) of Ref. \citep{0741-3335-54-3-035003} (also $\eta_{i}\equiv0$)
to obtain 
\begin{equation}
\begin{split} & e^{-i\frac{\pi}{4}}\sqrt{\frac{2\nu_{ei}}{\omega_{T}}}\frac{\delta_{*}}{\rho_{i}}\sqrt{1+1/\tau}\sqrt{1-\frac{\omega_{T}}{\omega}+4/\tau+2\sqrt{1/\tau(1+1/\tau)}}\times\\
 & \left\{ \frac{\Delta^{\prime}\rho_{i}}{\pi\hat{\beta}_{T}}-\frac{\omega^{2}}{\omega_{T}^{2}}\frac{1/\tau}{\sqrt{\pi}(1+1/\tau)^{2}}\ln\left(e^{i\frac{\pi}{4}}\frac{\rho_{i}}{\delta_{*}}\sqrt{\frac{\omega_{T}}{2\nu_{ei}}}\sqrt{\frac{\omega_{T}}{\omega}}\right)+\frac{\omega^{2}}{\omega_{T}^{2}}\bar{I}(\tau)\right\} +\\
 & \sqrt{\frac{\omega}{\omega_{T}}1/\tau}\left\{ 1-\frac{\Delta^{\prime}\rho_{i}\hat{\beta}_{T}}{\pi}\frac{\omega^{2}}{\omega_{T}^{2}}I(\tau)\right\} \left\{ \left(\frac{\omega}{\omega_{T}}-1\right)\sqrt{1+1/\tau}+\frac{\omega}{\omega_{T}}\sqrt{1/\tau}\right\} =0,
\end{split}
\label{eq:coll_DT}
\end{equation}
 where $\delta_{*}=\omega_{T}/(k_{y}v_{the})L_{s}$, and the charge
number $Z$ is set to unity for simplicity. This is equivalent to
evaluating the electron integral $I_{e}$ in Eq. (\ref{eq:disprelfinal})
using the electron conductivity defined in Eq. (\ref{eq:cond}).

\subsubsection{Cold ions limit $\tau\ll1$ and small $\Delta^{\prime}$}

We know that $\bar{I}\sim\tau^{1/2},$ for $\tau\ll1$ \citep{0741-3335-54-3-035003}.
We take this limit in Eq. (\ref{eq:coll_DT}). We also consider small
$\Delta^{\prime}$ for simplicity. Thus we have 
\begin{equation}
e^{-i\frac{\pi}{4}}\sqrt{\frac{2\nu_{ei}}{\omega_{T}}}\frac{\delta_{*}}{\rho_{i}}\sqrt{6}\frac{\Delta^{\prime}\rho_{i}}{\pi\hat{\beta}_{T}}=-\sqrt{\frac{\omega}{\omega_{T}}}\left(2\frac{\omega}{\omega_{T}}-1\right).\label{eq:DT_coll2}
\end{equation}
 For small, negative $\Delta^{\prime}=-2k_{y},$ Eq. (\ref{eq:DT_coll2})
gives a stable solution $\omega\approx\omega_{T}/2+(2/\pi)\sqrt{3}(k_{y}\delta_{*}/\hat{\beta}_{T})\sqrt{2\nu_{ei}/\omega_{T}}\exp[-i\pi/4].$
For small positive $\Delta^{\prime},$ after setting $\omega=\omega+i\gamma,$
with $\omega,\gamma\in\mathbb{R},$ we obtain two equations for the
real and imaginary parts of Eq. (\ref{eq:DT_coll2}) 
\begin{equation}
0=\frac{\omega}{\omega_{T}}\left[\left(2\frac{\omega}{\omega_{T}}-1\right)^{2}-4\frac{\gamma^{2}}{\omega_{T}^{2}}\right]-4\frac{\gamma^{2}}{\omega_{T}^{2}}\left(2\frac{\omega}{\omega_{T}}-1\right)\label{eq:re}
\end{equation}

\begin{equation}
-\frac{2\nu_{ei}}{\omega_{T}}\frac{\delta_{*}^{2}}{\rho_{i}^{2}}6\left(\frac{\Delta^{\prime}\rho_{i}}{\pi\hat{\beta}_{T}}\right)^{2}=\frac{\gamma}{\omega_{T}}\left[\left(2\frac{\omega}{\omega_{T}}-1\right)^{2}-4\frac{\gamma^{2}}{\omega_{T}^{2}}\right]+\frac{\omega}{\omega_{T}}4\frac{\gamma^{2}}{\omega_{T}^{2}}\left(2\frac{\omega}{\omega_{T}}-1\right).\label{eq:im}
\end{equation}
 We look for a solution $\omega=\omega_{T}\left(\frac{1}{2}+\epsilon\right).$
If $\frac{\gamma}{\omega_{T}}>1>\epsilon,$ then from Eq. (\ref{eq:im})
we obtain 
\begin{equation}
\left(\frac{\gamma}{\omega_{T}}\right)^{3}\approx\frac{\nu_{ei}}{\omega_{T}}\frac{\delta_{*}^{2}}{\rho_{i}^{2}}3\left(\frac{\Delta^{\prime}\rho_{i}}{\pi\hat{\beta}_{T}}\right)^{2},
\end{equation}
 whereas from Eq. (\ref{eq:re}) we find $\epsilon\approx-\frac{1}{6}.$
After converting the growth rate into Alfvénic units, we have 
\begin{equation}
\omega\approx\frac{1}{3}\omega_{T},
\end{equation}

\begin{equation}
\frac{\gamma}{\omega_{A}}\approx3^{1/3}\left(k_{y}L_{s}\right)^{2/3}S_{\eta}^{-1/3}\left(\frac{\Delta^{\prime}\rho_{s}}{2\pi}\right)^{2/3},\label{eq:forse}
\end{equation}
 with $\rho_{s}=\sqrt{1/(2\tau)}\rho_{i},$ and $S_{\eta}=v_{A}L_{s}/\eta$
is the Lundquist number. The growth rate is the same as Eq. (98B)
of Ref. \citep{zocco:102309}, but now the mode rotates with a frequency
$\omega\approx\frac{1}{3}\omega_{T}.$ This is the small $\Delta^{\prime}$
semicollisional drift-tearing mode \citep{drake:1341}. Notice that
we are solving for $\gamma$ real, hence Eq. (\ref{eq:forse}) is
not valid for negative $\Delta^{\prime}.$ 

When $\Delta^{\prime}\rho_{s}$ is large enough, we balance the two
terms multiplying $\Delta^{\prime}$ in Eq. (\ref{eq:coll_DT}). The
ion integral $I$, in the cold ion limit, can be calculated analytically
after using the Pad\'e approximant $(Z/\tau)(1-\hat{\Gamma}_{0})=-\rho_{s}^{2}\partial_{x}^{2}/[1-\rho_{i}^{2}\partial_{x}^{2}/2]$
for the ion response \citep{pegoraro:364}; then we obtain 
\begin{equation}
\left(\frac{\omega}{\omega_{T}}\right)^{5/2}\left(2\frac{\omega}{\omega_{T}}-1\right)\approx\frac{2}{\pi}\sqrt{3}e^{-i\frac{\pi}{4}}\sqrt{2\frac{\nu_{ei}}{\omega_{T}}}\frac{\delta_{*}}{\rho_{s}}\frac{1}{\hat{\beta}_{T}^{2}}.
\end{equation}
 When $\omega/\omega_{T}\gg1,$ we find 
\begin{equation}
\left|\frac{\omega}{\omega_{T}}\right|\approx\left(\frac{\sqrt{3}}{\pi}\sqrt{2\frac{\nu_{ei}}{\omega_{T}}}\frac{\delta_{*}}{\rho_{s}}\frac{1}{\hat{\beta}_{T}^{2}}\right)^{2/7}.\label{eq:coppimio}
\end{equation}
 This result will be confirmed by the numerical solution of Eq. (\ref{eq:coll_DT}).

\subsubsection{Solution of Equation (\ref{eq:coll_DT})}

We now solve the dispersion relation (\ref{eq:coll_DT}) numerically
for arbitrary positive $\Delta^{\prime}.$ For this we choose $\rho_{e}/L_{T}=10^{-3},$
$1/\tau=100,$ $\nu_{ei}/\omega_{T}=18,$ $d_{e}/\rho_{s}=\sqrt{2}\times0.08,$
and $k_{y}L_{s}=2.$ In this way $\delta_{0}/\rho_{i}\equiv\sqrt{2\nu_{ei}/\omega_{T}}\delta_{*}/\rho_{i}=8.5\times10^{-4}.$
The electron inertia $d_{e}$ is neglected in Ohm's law, but defines
$\hat{\beta}^{2}=2\delta_{*}/d_{e},$ with $\delta_{*}=\omega_{T}/(k_{y}v_{the})L_{s}$
\citep{zocco:102309}. The result is shown in Fig. (3). The analytical
result is reproduced very well. We can also notice that, for $\Delta^{\prime}\rho_{s}\gg1$,
the growth rate does not depend on $\Delta^{\prime}$ \citep{coppi-res},
and agrees with Eq. (\ref{eq:coppimio}). For $\Delta^{\prime}\rho_{s}\ll1$
a diamagnetic stabilisation occurs, $\gamma<\omega,$ and Eq. (64)-(65)
are no longer valid. 
\begin{figure}[!h]
\input{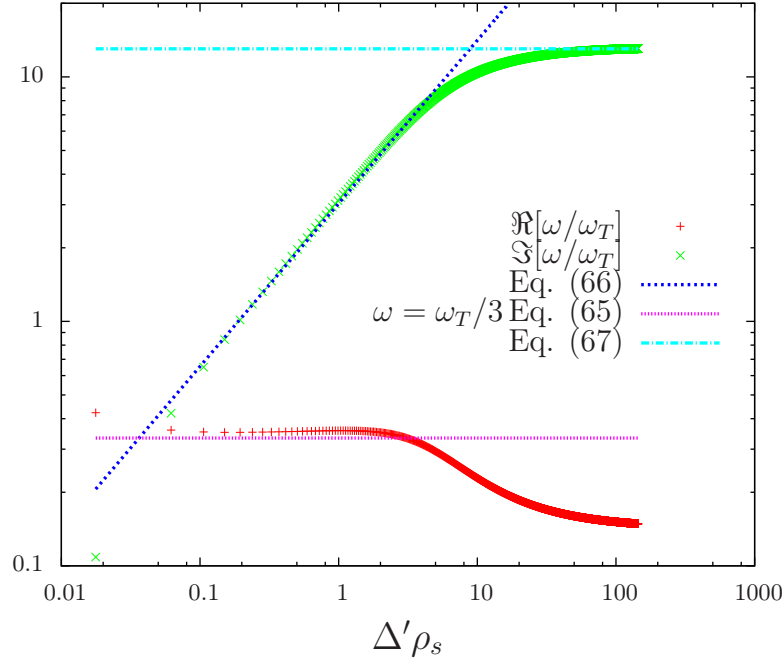} 

\caption{The real and imaginary part of the numerical solution of Eq. (\ref{eq:coll_DT})
as a function of $\Delta^{\prime}\rho_{s}.$ Here, $\rho_{e}/L_{T}=10^{-3},$
$1/\tau=100,$ $\nu_{ei}/\omega_{T}=18,$ $d_{e}/\rho_{s}=\sqrt{2}\times0.08,$
$k_{y}L_{s}=2$ and $\delta_{0}/\rho_{i}=8.5\times10^{-4}.$ The lines
are from the analytical solutions in Eqs. (64), (65) and (67).}

\label{fig:numplot} 
\end{figure}


\section{Weakly collisional limit\label{sec:Weakly-collisional-limit}}

\subsection{Fluid Limit}

\begin{figure}[!h]
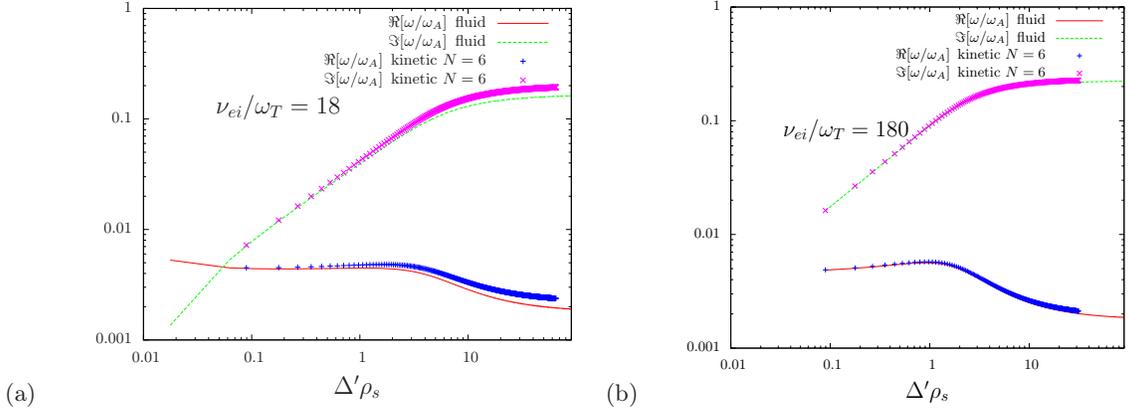

\subfloat{(a)}{\scalebox{.65}{\input{collplot2.tex}}}\subfloat{(b)}{\scalebox{.6}{\input{collplot3.tex}}} 

\caption{The numerical solution of Eq. (\ref{eq:coll_DT}) (solid lines) and
of Eq. (\ref{eq:disprelfinal}) (with $N=6$) (symbols) for $\rho_{e}/L_{T}=10^{-3},$
$1/\tau=100,$ $d_{e}/\rho_{s}=\sqrt{2}\times0.08,$ $k_{y}L_{s}=2$
and $\delta_{0}/\rho_{i}=8.5\times10^{-4}.$ Here $\nu_{ei}/\omega_{T}=18,$
(a) and $\nu_{ei}/\omega_{T}=180,$ (b). The fluid limit (solid lines)
is reproduced. }
\label{fig:collplot-1-1} 
\end{figure}

Before studying the microtearing mode, let us verify that Eq. (\ref{eq:disprelfinal}),
in the limit $\nu_{ei}/\omega\gg1$, agrees with the drift-tearing
dispersion relation Eq. (\ref{eq:coll_DT}), which has been derived
analytically. The results are shown in Fig. (\ref{fig:collplot-1-1})
(a) and (b). Here the solid lines represent the solution of Eq. (\ref{eq:coll_DT})
for $\nu_{ei}/\omega_{T}=18$ and $\nu_{ei}/\omega_{T}=180$, where
the electronic integral $I_{e}$ defined in Eq. (\ref{eq:Ie}) was
performed analytically using the semicollisional conductivity in Eq.
(\ref{eq:cond}). Other parameters are as in Fig. (\ref{fig:numplot})
above. The symbols in Fig. (\ref{fig:collplot-1-1}) represent the
solution calculated using Eq. (\ref{eq:disprelfinal}), with $N=6$
Hermite moments. For this velocity space resolution, the collisional
non-isothermal (fluid) limit is recovered very well and the agreement
improves with higher collisionality.

\subsection{Weakly Collisional Drift-tearing mode}

\begin{figure}[!h]
\input{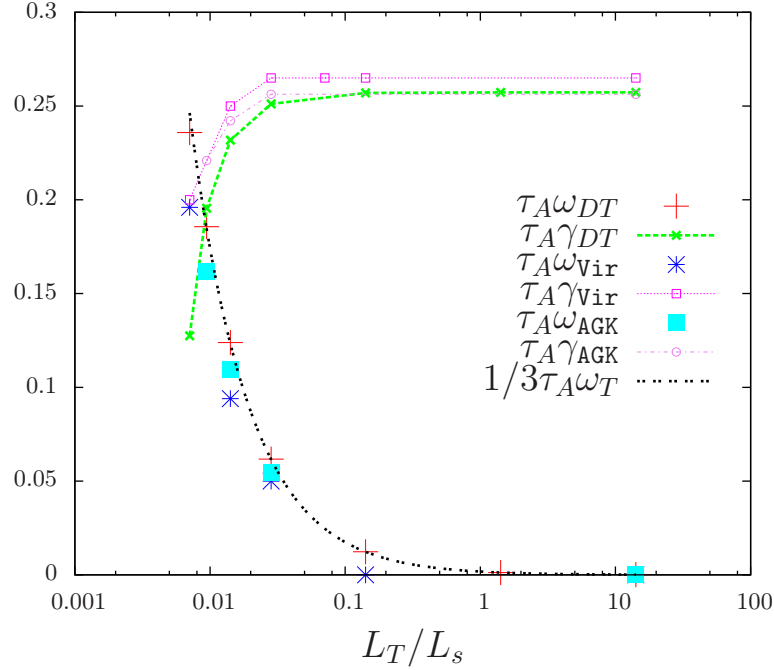} 

\caption{The numerical solution of Eq. (\ref{eq:disprelfinal}) ($N=20$) compared
to the solutions obtained by the {\tt Viriato} code and {\tt AstroGK}
code. Here $d_{e}/L_{s}=\rho_{s}/L_{s}=0.2,$ $\Delta^{\prime}L_{s}=23.2/\sqrt{2},$
$\nu_{ei}/(v_{A}/L_{s})=.02/\sqrt{2},$ $\tau=1,$ and $\tau_{A}\equiv\omega_{A}^{-1}.$
Results are presented in Alfv\'enic units. The subscript $DT$ stands
for ``drift-tearing''.}
\label{fig:virAGKbench-1} 
\end{figure}

In the weakly collisional limit, we compare the solution of Eq. (\ref{eq:disprelfinal})
with the results of the hybrid fluid-kinetic {\tt Viriato} code that
solves Eqs. (\ref{eq:elcont}), (\ref{eq:grnohmslaw}) and (\ref{eq:sum3})
\citep{loureiro-scecco-zocco}, and the gyrokinetic code {\tt AstroGK}
(AGK)\citep{astroGK}. The level of agreement is satisfactory, see
Fig. (\ref{fig:virAGKbench-1}).

\newpage{}

\newpage

\section{Microtearing Mode\label{sec:Microtearing-Mode}}

Now we turn our attention to the microtearing mode, meaning that we
analyse the case in which $\Delta^{\prime}<0.$ We start with one
example that relates this work to previous theories, thus allowing
a comparative approach. Following Gladd et al. \citep{Gladdmicrotearing},
we neglect the electrostatic potential in Eqs. (\ref{eq:semi1}) and
(\ref{eq:semi2}) and obtain 
\begin{equation}
\left(\frac{d^{2}}{dx^{2}}-k_{y}^{2}\right)A_{\parallel}\approx-i\frac{\omega}{\nu_{ei}}\frac{1}{d_{e}^{2}}\frac{1}{1-i\frac{\omega}{\nu_{ei}}}\hat{\sigma}_{e}A_{\parallel}.
\end{equation}
 In the case of constant-$A_{\parallel},$ this equation is matched
to ideal MHD in the usual way, hence \citep{Gladdmicrotearing} 
\begin{equation}
\Delta^{\prime}d_{e}=-i\frac{\omega}{\nu_{ei}}\frac{2}{d_{e}}\frac{1}{1-i\frac{\omega}{\nu_{ei}}}\int_{0}^{\infty}dx\hat{\sigma}_{e}(x/\delta).\label{eq:gladdmicro}
\end{equation}
The integral dispersion relation (\ref{eq:gladdmicro}) can be studied
using different electron conductivity models, and performing several
subsidiary expansions; in particular for small $\omega/\nu_{ei}$
\citep{Gladdmicrotearing}. However, in this limit, the growth rate
is always a subdominant correction to the stabilising term which is
proportional to $-\left|\Delta\right|^{\prime}d_{e}.$ The real frequency
of the mode can also be found is some subsidiary expansion. For instance,
we could consider the limit $\Delta^{\text{\ensuremath{\prime}}}d_{e}\ll1,$
in analogy to our treatment of Eq. (\ref{eq:coll_DT}). Yet, the analogy
between Eq. (\ref{eq:gladdmicro}) and (\ref{eq:coll_DT}), or the
more general (\ref{eq:disprelfinal}), is not only formal, as we are
about to show.

Equation (\ref{eq:gladdmicro}) has been derived for unmagnetised
ions, and neglecting the electrostatic potential in Ohm's law. Under
these circumstances, one can easily convince oneself that the microtearing
theory derived using Eq. (\ref{eq:gladdmicro}) is a simplified version
of the low-$\hat{\beta}_{T}$ and low-$\delta k_{y}$ drift-tearing
theory just derived in Section (\ref{sec:Dispersion-Relation}). In
previous works \citep{antonsen-coppi,coppi-collisionless,PhysRevLett.42.1058},
the fundamental frequency of the drift-tearing mode, $\omega_{DT}/\omega_{T},$
solution of $I_{e}(\omega_{DT}/\omega_{T})=0$, has been derived by
neglecting the electrostatic potential, therefore replacing
\[
\begin{split}I_{e} & =-\int_{0}^{\infty}ds\frac{F_{\infty}\hat{\sigma}_{e}(s)}{\left(1-i\frac{\omega}{\nu_{ei}}\right)F_{\infty}-s^{2}\hat{\sigma}_{e}(s)}\rightarrow-\frac{1}{\left(1-i\frac{\hat{\omega}}{\hat{\nu}_{ei}}\right)}\int_{0}^{\infty}ds\hat{\sigma}_{e}(s),\end{split}
\]
which is exactly the same approximation used to derive Eq. (\ref{eq:gladdmicro}).
Thus, according to these results, for $\Delta^{\prime}=0$, to leading
order, one expects a marginally stable collisionless drift-tearing
mode \citep{antonsen-coppi,coppi-collisionless,PhysRevLett.42.1058}.
Since both drift-tearing and microtearing modes must be derived from
the same equation when the electrostatic potential is not neglected,
we infer that, if there were any low$-\hat{\beta}_{T}$ microtearing
mode, this must also be marginally stable for negligible collisionality.
Furthermore, if there were any instability arising from Eq. (\ref{eq:gladdmicro}),
a critical $\hat{\beta}_{T}$ for instability could be calculated.
However, that would be incorrect, since in Eq. (\ref{eq:gladdmicro})
the electrostatic potential was neglected. The parameter $\hat{\beta}_{T}$
is in fact a measure of the ratio $A_{\parallel}/\varphi.$ If the
electrostatic potential is neglected with impunity, from Eq. (\ref{eq:gladdmicro})
it might seem possible to reach arbitrarily large values of $\hat{\beta}_{T}.$
We can understand why this is not appropriate, and a new high-$\hat{\beta}_{T}$
theory is thus required. Firstly, the microtearing is an electromagnetic
mode driven by the electron temperature gradient; hence, increasing
$\hat{\beta}_{T}$ by decreasing $L_{T}$ should enhance the instability.
Therefore, we expect the growth rate to be subdominant in a low-$\hat{\beta}_{T}$
theory. Secondly, from Eq. (\ref{eq:disprelfinal}) we notice the
following. On the LHS we have the tearing mode driving term (stabilizing
for $\Delta^{\prime}\text{\ensuremath{\rho}}_{i}<0$). If one replaces
$\Delta^{\prime}=-2k_{y},$ and approximates $\omega\approx\omega_{T}$
in the definition of $\delta,$ it immediately becomes clear that
the whole LHS is proportional to the inverse power of $\hat{\beta}_{T}.$
Therefore, stabilising terms can be subdominant for high values of
$\hat{\beta}_{T}.$ However, the higher the $\hat{\beta}_{T},$ the
more important the non-constant-$A_{\parallel}$ contribution on the
RHS of Eq. (\ref{eq:disprelfinal}) (the $C$ term). This fact explains
why, in microtearing simulations \citep{Gladdmicrotearing}, the constant-$A_{\parallel}$
approximation was found to be violated, and invalidates analytical
theories that relied on this approximation. We conclude that, at high
$\hat{\beta}_{T}$ (necessary for instability) the constant-$A_{\parallel}$
approximation cannot be used.

In order to capture these aspects, a new high-$\hat{\beta}_{T}$ theory
is needed. Nevertheless, before embarking on this task, we find it
useful to artificially suppress all the stabilising terms on the LHS
of Eq. (\ref{eq:disprelfinal}) (which was derived in a low-$\hat{\beta}_{T}$
expansion), and solve for $I_{e}=0$ for kinetic electrons and finite
collisionality. Indeed, we are expecting to observe a microtearing
mode with a growth rate which is a nonmonotonic function of the collision
frequency. Even if Eq. (\ref{eq:disprelfinal}) is not enough to describe
an unstable microtearing mode, the solution of $I_{e}=0$ can still
give an eigenvalue with an imaginary part which is a nonmonotonic
function of collisionality when the tearing mode is marginally stable,
that is when $\Delta^{\prime}$ reaches values that are solutions of
the equation $B=0.$ We could therefore study the properties of this
maximum, which would eventually give an instability for large enough
$\hat{\beta}_{T}.$ In other words: Eq. (\ref{eq:disprelfinal}) surely
captures some salient features of a \textsl{stable }microtearing mode,
but we need a new theory to describe an unstable one in a consistent
way.

\begin{figure}[!h]
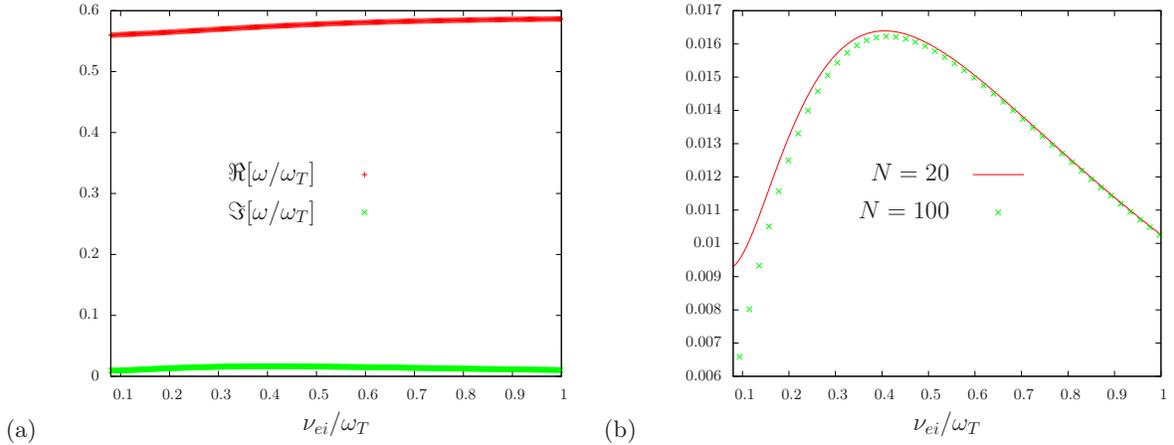

\subfloat{(a)}{\scalebox{0.65}{\input{IedeyesRE.tex}}} \subfloat{(b)}{\scalebox{0.65}{\input{IedeyesIM.tex}}} 

\caption{The solution of $I_{e}=0.$ Real and imaginary part (a), and a zoom
of the imaginary part (b). }
\label{fig:Ie}
\end{figure}

We then solve for $I_{e}=0.$ Results are shown in Fig. (\ref{fig:Ie}).
The real frequency is close to the familiar value $\omega\approx0.5\omega_{T}$
\citep{antonsen-coppi,coppi-collisionless,PhysRevLett.42.1058}, and
the ``growth rate'' is indeed a non-monotonic function of collisionality
{[}see Fig. (\ref{fig:Ie}){]}. This non-monotonic dependence of the
imaginary part of the eigenvalue with collisionality remains the invariant
feature of the mode in the literature \citep{PhysRevLett.106.155004,doerk:055907,PhysRevLett.108.235002,PhysRevLett.110.155005}.
\begin{figure}[!h]
\input{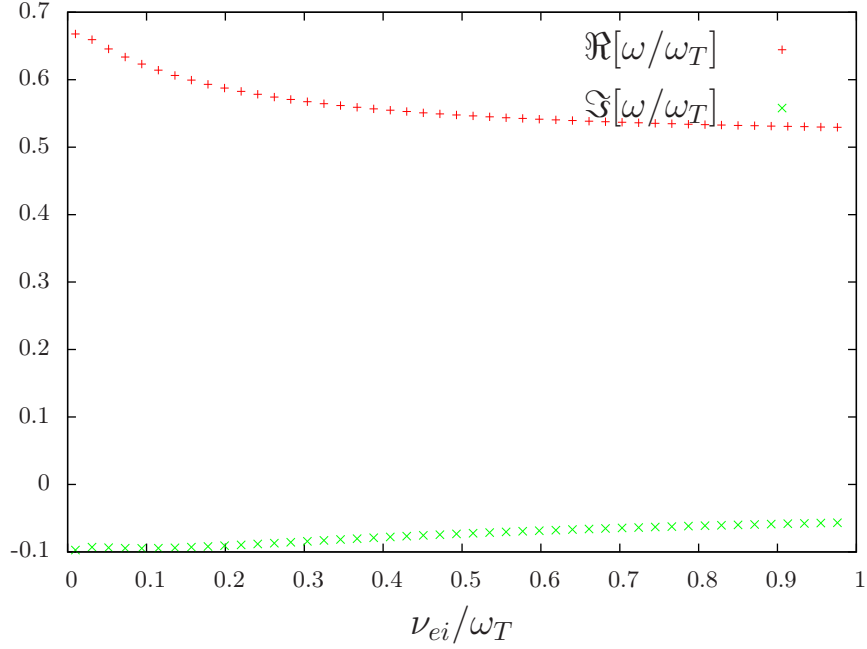}

\caption{Solution of $I_{e}=0.$ Real and imaginary part of the eigenvalue
for $d_{e}\equiv0$ in Ohm's law. Here $N=100.$ The mode is always
stable.}

\label{fig:node}
\end{figure}
We verified that the instability is present only if we include electron
inertia {[}see Fig (\ref{fig:node}){]}, the electrostatic potential
and if we consider non-isothermal electrons, that is $N>2.$ Electron
inertia is neglected when the integral $I_{e}$ is replaced with 
\[
I_{e}\rightarrow-\int_{0}^{\infty}ds\frac{F_{\infty}\tilde{\sigma}}{F_{\infty}-s^{2}\tilde{\sigma}},
\]
where
\begin{equation}
\tilde{\sigma}=\frac{\sigma_{0}+\frac{3\nu_{ei}}{-i\omega+3\nu_{ei}+\frac{4}{2}\frac{k_{\parallel}^{2}v_{the}^{2}}{\Omega(N)}}s^{2}}{1+\left[3+\frac{3\nu_{ei}}{-i\omega+3\nu_{ei}+\frac{4}{2}\frac{k_{\parallel}^{2}v_{the}^{2}}{\Omega(N)}}\right]s^{2}+\frac{3\nu_{ei}}{-i\omega+3\nu_{ei}+\frac{4}{2}\frac{k_{\parallel}^{2}v_{the}^{2}}{\Omega(N)}}s^{4}}.
\end{equation}
Electron inertia is also required to obtain a frequency that tends
to $\omega\approx0.5\omega_{T}$ in the weakly collisional limit {[}see
Fig. (\ref{fig:Ie}){]}. While the nonmonotonic shape of the growth
rate is determined by electron kinetics, the coupling to the kinetic
Alfv\'en wave {[}$A_{\parallel}$non-constant, the second term in
Eq. (\ref{Peg}){]} determines whether the peak will reach positive
values. This is the subject of the next Section.

\subsection{High-$\hat{\beta}_{T}$ theory}

In the previous section, we identified the reason for obtaining a
low-$\hat{\beta}_{T}$ mode with an imaginary part which is a nonmonotonic
function of collisionality. To have microtearing instability, high
$\hat{\beta}_{T}$'s are needed to overcome the stabilising effect
of a negative $\Delta^{\prime}.$ As $\hat{\text{\ensuremath{\beta}}_{T}}$
is increased, breaking of the constant-$A_{\parallel}$ approximation
occurs, and the term $C$ in Eq. (\ref{eq:disprelfinal}) is no longer
unity. Therefore, the low-$\hat{\beta}_{T}$ theory described by Eq.
(\ref{eq:disprelfinal}) is not sufficient to describe an unstable
microtearing mode, even if it captures some salient features. A high-$\hat{\beta}_{T}$
theory, with non-constant-$A_{\parallel},$ is a straightforward modification
of that presented by Connor et al. \citep{0741-3335-54-3-035003}.
This theory resembles a previous theory formulated by Drake et al.
\citep{drake-phfl-84-rmhd}. However, in their work, Connor et al.
proved that a high-$\hat{\beta}_{T}$ theory matches exactly onto
a low-$\hat{\beta}_{T}$ one, as in Eq. (\ref{eq:disprelfinal}),
when an appropriate ''screening factor'' is taken into account.

We present such a high-$\hat{\beta}_{T}$ theory, and give an explicit
analytic expression for the growth rate of the microtearing mode that
matches the low-$\hat{\beta}_{T}$ theory. We rewrite Eq. (\ref{eq:conduttivita})
in the following way
\begin{equation}
\hat{\sigma}_{e}=\frac{\text{\ensuremath{\sigma}}_{0}+\sigma_{1}s^{2}}{1+d_{0}s^{2}+d_{1}s^{4}},
\end{equation}
with
\begin{equation}
\sigma_{0}=1-\hat{\text{\ensuremath{\omega}}}^{-1},
\end{equation}
\begin{equation}
\sigma_{1}=\frac{3\nu_{ei}}{-i\omega+3\nu_{ei}+\frac{4}{2}\frac{k_{\parallel}^{2}v_{the}^{2}}{\Omega(N)}},
\end{equation}
\begin{equation}
d_{0}=\left[\frac{3}{1-i\frac{\omega}{\nu_{ei}}}+\frac{3\nu_{ei}}{-i\omega+3\nu_{ei}+\frac{4}{2}\frac{k_{\parallel}^{2}v_{the}^{2}}{\Omega(N)}}\right],
\end{equation}
and
\begin{equation}
d_{1}=\frac{1}{1-i\frac{\omega}{\nu_{ei}}}\frac{3\nu_{ei}}{-i\omega+3\nu_{ei}+\frac{4}{2}\frac{k_{\parallel}^{2}v_{the}^{2}}{\Omega(N)}}.
\end{equation}
We first consider the $\text{\ensuremath{\hat{\text{\ensuremath{\beta}}}}}_{T}\gg1$
limit, with $\hat{\text{\ensuremath{\omega}}}^{2}\text{\ensuremath{\sim}}\text{\ensuremath{\hat{\text{\ensuremath{\beta}}}}}_{T}^{-1}\ll1,$
so that $\hat{\text{\ensuremath{\omega}}}^{2}\text{\ensuremath{\hat{\text{\ensuremath{\beta}}}}}_{T}\sim\mathcal{O}(1).$
In this limit, we expect diamagnetic effects to screen the resonant
layer, thus preventing reconnection. When approaching the ion region,
for $s=x/\delta\gg1,$ the equation for the current in the electron
region {[}Eq. (\ref{eq:curr}){]} becomes
\begin{equation}
\frac{d^{2}}{ds^{2}}\bar{J}=\hat{\text{\ensuremath{\omega}}}^{2}\text{\ensuremath{\hat{\text{\ensuremath{\beta}}}}}_{T}\frac{F_{\infty}}{F_{\infty}-1}\frac{\sigma_{0}+\sigma_{1}^{\infty}s^{2}}{\sigma_{1}^{\infty}s^{4}}\bar{J},\label{eq:currshield}
\end{equation}
with
\begin{equation}
\bar{J}=\frac{1-i\frac{\omega}{\nu_{ei}}+\left[\left(1-i\frac{\omega}{\nu_{ei}}\right)d_{0}-\frac{\sigma_{0}}{F_{\infty}}\right]s^{2}+\frac{F_{\infty}-1}{F_{\infty}}\sigma_{1}^{\infty}s^{4}}{\sigma_{0}+\sigma_{1}^{\infty}s^{2}}J,\label{eq:currsielsol}
\end{equation}
and 
\begin{equation}
\sigma_{1}^{\infty}=\lim_{s\rightarrow\infty}\sigma_{1}.
\end{equation}
We do not need to calculate explicitly the ``screening factor''
here. We choose the solution of Eq. (\ref{eq:currshield}) that is
small (completely screened by diamagnetic effects) at $s=0$ \citep{drake:2509,0741-3335-54-3-035003}
\begin{equation}
J=\frac{\sigma_{0}+\sigma_{1}s^{2}}{1-i\frac{\omega}{\nu_{ei}}+\left[\left(1-i\frac{\omega}{\nu_{ei}}\right)d_{0}-\frac{\sigma_{0}}{F_{\infty}}\right]s^{2}+\frac{F_{\infty}-1}{F_{\infty}}\sigma_{1}s^{4}}\sqrt{\frac{s}{s_{t}}}K_{\mu}\left(\frac{s_{t}}{s}\right),\label{eq:currscreen}
\end{equation}
where $K_{\mu}$ is the modified Bessel function, 
\begin{equation}
s_{t}^{2}=\left(\mu^{2}-\frac{1}{4}\right)\frac{\sigma_{0}}{\sigma_{1}^{\infty}}=\hat{\beta}_{T}\hat{\omega}\frac{Z/\tau}{Z/\tau+1},\text{\,\,\mbox{for}\,}\nu_{ei}\gg\omega,
\end{equation}
and $\mu$ is defined as in the low-$\hat{\beta}_{T}$ case, $1/4-\mu^{2}=\hat{\omega}^{2}\hat{\beta}_{T}F_{\infty}/(F_{\infty}-1),$
but we do not approximate $\mu\neq1/2+\delta\text{\ensuremath{\mu}},$
with $\delta\mu\ll1,$ unlike in the low-$\hat{\beta}_{T}$ case.
Solution (\ref{eq:currscreen}) has to be matched to the ion region
solution. In the high-$\hat{\beta}_{T}$ regime, a Pad\'e approximant
\citep{pegoraro:364,0741-3335-54-3-035003} is adequate to describe
the ion response \citep{0741-3335-54-3-035003}. Then, the coefficients
$\hat{a}_{\pm}$ in Eq. (\ref{eq:gendisprel}) are \citep{0741-3335-54-3-035003}
\begin{equation}
\frac{\hat{a}_{-}}{\hat{a}_{+}}=\left[\frac{1}{2}\left(1+\frac{Z}{\tau}\right)\right]^{-\mu}\frac{\Gamma(-\mu)}{\Gamma(\mu)}\frac{\frac{1}{\Gamma^{2}\left(-\frac{\text{\ensuremath{\mu}}}{2}-\frac{1}{4}\right)}-\frac{\pi\sqrt{2}}{8}\frac{\hat{\beta}_{T}}{\Delta^{\prime}\rho_{\tau}}\frac{1}{\Gamma^{2}\left(-\frac{\text{\ensuremath{\mu}}}{2}-\frac{5}{4}\right)}}{\frac{1}{\Gamma^{2}\left(\frac{\text{\ensuremath{\mu}}}{2}-\frac{1}{4}\right)}-\frac{\pi\sqrt{2}}{8}\frac{\hat{\beta}_{T}}{\Delta^{\prime}\rho_{\tau}}\frac{1}{\Gamma^{2}\left(\frac{\text{\ensuremath{\mu}}}{2}-\frac{5}{4}\right)}},
\end{equation}
with $\rho_{\tau}=\sqrt{\frac{1}{2}\left(1+\frac{Z}{\tau}\right)}\rho_{i}.$
After using Eq. (\ref{eq:formula}), with $b_{\pm}$ calculated using
the large asymptotic expansion of solution (\ref{eq:currscreen}),
we obtain the high-$\hat{\beta}_{T}$ dispersion relation$^{1}$\footnotetext[1]{After
some algebra, it is easy to show that this dispersion relation is
the equivalent of Eq. (11) of Pegoraro et al. \citep{pegoraro:364}.
We keep it in the form of Ref. \citep{drake:2509} to stress the fact
that the Fourier space analysis of the ion region  \citep{pegoraro:364,0741-3335-54-3-035003}
gives the same result as the real space analysis. } 
\begin{equation}
e^{i\frac{\pi}{2}\text{\ensuremath{\mu}}}\left(\frac{2\sigma_{1}^{\infty}}{\frac{1}{4}-\mu^{2}}\frac{\rho_{\tau}^{2}}{\delta_{0}^{2}}\right)^{\mu}=\frac{\mu+\frac{1}{2}}{-\mu+\frac{1}{2}}\frac{\Gamma^{2}(-\mu)}{\Gamma^{2}(\mu)}\frac{\mathcal{D}-\cot\left[\pi\left(\frac{1}{4}+\frac{\text{\ensuremath{\mu}}}{2}\right)\right]}{\mathcal{D}-\cot\left[\pi\left(\frac{1}{4}-\frac{\text{\ensuremath{\mu}}}{2}\right)\right]},\label{eq:disphighb}
\end{equation}
where 
\begin{equation}
\mathcal{D}=\frac{2}{\text{\ensuremath{\pi}}}\Delta^{\prime}\rho_{\tau}\frac{\Gamma\left(\frac{5}{4}-\frac{\text{\ensuremath{\mu}}}{2}\right)\Gamma\left(\frac{5}{4}+\frac{\text{\ensuremath{\mu}}}{2}\right)}{\Gamma\left(\frac{3}{4}-\frac{\text{\ensuremath{\mu}}}{2}\right)\Gamma\left(\frac{3}{4}+\frac{\text{\ensuremath{\mu}}}{2}\right)}.
\end{equation}
This is the large $\eta_{e}\gg1$ ($\omega_{*e}\ll1$) limit of Eq.
(81) of Ref. \citep{0741-3335-54-3-035003} for kinetic electrons,
when the screening factor is set to unity. In the collisional limit,
$\nu_{ei}\gg\omega,$ the low $\hat{\beta}_{T}$ limit of Eq. (\ref{eq:disphighb})
connects to the low $\hat{\omega}^{2}\ll1$ limit of Eq. (\ref{eq:disprelfinal}).
Thus, from Eq. (\ref{eq:coll_DT}), using $\Delta^{\prime}\rightarrow-2k_{y},$
and $\hat{\omega}^{2}\ll1,$ we have
\begin{equation}
e^{i\frac{\pi}{4}}\sqrt{2\frac{\tau}{Z}\frac{\nu_{ei}}{\omega_{T}}}\frac{\delta_{*}}{\hat{\beta}_{T}^{2}\rho_{\tau}}\frac{2k_{y}\rho_{\tau}\hat{\beta}_{T}}{\pi}+\frac{\omega}{\omega_{T}}\left(1+k_{y}\rho_{\tau}\hat{\beta}_{T}\frac{1/\tau+1}{1/\tau}\frac{\omega^{2}}{\omega_{T}^{2}}\right)=0.\label{eq:MTMT}
\end{equation}
When the constant-$A_{\parallel}$ approximation fails, $C>1$ in
Eq. (\ref{eq:disprelfinal}), we have 
\begin{equation}
\hat{\beta}_{T}>\frac{1}{1+\tau}\frac{\pi}{2}\frac{1}{k_{y}\rho_{\tau}\omega^{2}/\omega_{T}^{2}},
\end{equation}
therefore, for 
\begin{equation}
\frac{2}{\pi}\left(k_{y}\rho_{\tau}\right)^{2/3}\left(\frac{d_{e}}{\rho_{\tau}}\right)^{1/3}\left(\frac{2\tau}{\sqrt{\hat{\beta}_{T}}}\frac{\nu_{ei}}{v_{the}/L_{s}}\right)^{1/3}>1,\label{eq:constpsi}
\end{equation}
and $\gamma^{2}>\omega_{0}^{2},$ with $\omega=\omega_{0}+i\gamma,$
we find an unstable mode
\begin{equation}
\omega=\text{\ensuremath{\omega}}_{T}\left(\frac{1}{3}+i\right)\left(2\tau\frac{\nu_{ei}}{\omega_{T}}\right)^{1/6}\left(\frac{1}{1+\tau}\right)^{1/3}\left(\frac{\delta_{*}}{\hat{\beta}_{T}^{2}\rho_{\text{\ensuremath{\tau}}}}\right)^{1/3},\label{MTcoll}
\end{equation}
the large-$\eta_{e}$ semicollisional microtearing mode. When electron
inertia is the relevant electron scale, we replace $\nu_{ei}$ by
$-i\omega$ in Eq. (\ref{MTcoll}), to obtain
\begin{equation}
\omega=\omega_{T}\left(2\tau\right)^{1/5}\left(\frac{1}{3}+i\right)^{6/5}e^{i\frac{\pi}{5}}\left(\frac{1}{1+\tau}\right)^{2/5}\left(\frac{\delta_{*}}{\hat{\beta}_{T}^{2}\rho_{\text{\ensuremath{\tau}}}}\right)^{2/5},\label{eq:MTcollless}
\end{equation}
the ``weakly-collisional'' large-$\eta_{e}$ microtearing mode.
Notice that these solutions are based on the expansion in $\epsilon\sim\omega_{0}^{2}/\gamma^{2}\sim1/9.$

A short digression on the finite $\eta_{e}$ theory of Connor et al.
\citep{0741-3335-54-3-035003} is now required. Equation (\ref{eq:MTMT})
is the equivalent of Eq. (45) of Ref. \citep{0741-3335-54-3-035003}.
In the large $\hat{\beta}_{N}=\hat{\beta}_{T}L_{T}^{2}/L_{n_{0e}}^{2}$
limit, the real frequency of the drift tearing mode, $\omega\approx\omega_{*e}(1+0.7\eta_{e}),$
is driven to $\omega\approx\omega_{*e}.$ More precisely, for finite
$\eta_{e},$ when $\omega\approx\omega_{*e},$ equation (\ref{eq:MTMT})
gives the solution with real frequency
\begin{equation}
\frac{\omega}{\omega_{*e}}=1+\frac{2}{1+\frac{Z}{\tau}}\left(\frac{1}{k_{y}\rho_{i}\hat{\beta}_{N}}\right)^{2}+\frac{8\sqrt{2}\sqrt{d_{0,c}+2\sqrt{d_{1,c}}-1.71\eta_{e}/\left(1+Z/\tau\right)}}{1.71\eta_{e}/\left(1+Z/\tau\right)}\frac{1}{k_{y}\rho_{i}\hat{\beta}_{N}^{3}}\frac{\delta_{N}}{\rho_{i}},\label{eq:omCHZ}
\end{equation}
and growth rate
\begin{equation}
\begin{split} & \frac{\gamma}{\omega_{*e}}=-\frac{16}{\sqrt{2}\pi}\frac{\sqrt{d_{0,c}+2\sqrt{d_{1,c}}-1.71\eta_{e}/\left(1+Z/\tau\right)}}{1.71\eta_{e}/\left(1+Z/\tau\right)}\frac{1}{k_{y}\rho_{i}\hat{\beta}_{N}^{3}}\frac{\delta_{N}}{\rho_{i}}\times\\
 & \left[1+\frac{1}{\sqrt{2}\pi}\frac{\delta_{N}}{\rho_{i}}\frac{k_{y}\rho_{i}}{\hat{\beta}_{N}}\frac{\sqrt{d_{0,c}+2\sqrt{d_{1,c}}-1.71\eta_{e}/\left(1+Z/\tau\right)}}{1.71\eta_{e}/\left(1+Z/\tau\right)}\right].
\end{split}
\label{eq:gammaCHZ}
\end{equation}
Here $\delta_{N}=\delta(\omega=\omega_{*e}),$ while $d_{0,c}$ and
$d_{1,c}$ are numerical factors that depend on the model collisional
operator. They are $d_{0,c}=5.08,$ and $d_{1,c}=2.13$ in the Braginski
collisional model \citep{drake:2509,cowley:3230,0741-3335-54-3-035003},
and $d_{0,c}=4,$ and $d_{1,c}=1$ in our present model. Thus, the
growth rate derived in Eq. (\ref{MTcoll}) is equivalent to Eq. (\ref{eq:gammaCHZ})
in the large $\eta_{e}$ limit, and the critical $k_{y}\rho_{\tau}$
in Eq. (\ref{eq:constpsi}) defines the value above which the constant-$A_{\parallel}$
approximation breaks down. This is the necessary condition for instability
in the flat density limit. When both $\omega_{*e}$ and $\eta_{e}$
are finite, the root in Eq. (\ref{eq:omCHZ}) is driven unstable for
\begin{equation}
\eta_{e}>\eta_{e}^{MT}=\frac{d_{0,c}+2\sqrt{d_{1,c}}}{1.71/\left(1+Z/\tau\right)}.\label{eq:crieta}
\end{equation}
We call this the finite-$\eta_{e}$ microtearing mode. The mode is
destabilised by electron temperature gradients when the parameter
$\hat{\beta}_{N}$ or $\hat{\beta}_{T}$ are sufficiently large to
break the constant-$A_{\parallel}$ approximation. For finite density
gradients, the mode is destabilised when $\eta_{e}>\eta_{e}^{MT},$
but an increasingly large temperature gradient has a \textsl{stabilising
}effect, since the growth rate scales like $\gamma\sim\omega_{*e}\eta_{e}^{-1/2}$.
When density gradients are negligible compared to temperature gradients,
the residual mode of Eq. (\ref{MTcoll}) remains. 

In this analysis, an energy dependent collision frequency is not required.
We show why this is the case. Let us rewrite the electron drift-kinetic
Eq. (\ref{eq:edk}) for $\delta f_{e}=h_{e}+e\varphi/T_{0e},$ and
use a Lorentz collision operator $\nu\partial_{\xi}(1-\xi^{2})\partial_{\xi},$
where $\xi=v_{\parallel}/v,$ and $\nu=\nu_{ei}\hat{v}^{-3\alpha}$
is the energy-dependent collision frequency. The equation can then
be solved by using an expansion in Legendre polynomials \citep{drake:1341}.
After truncating to first order, calculating the parallel electron
current, and using Ampere's law, in the $\nu_{ei}/\omega\gg1$ limit
one obtains the following electron conductivity
\begin{equation}
\sigma_{DL}\propto\int_{0}^{\infty}d\hat{v}\hat{v}^{4+3\alpha}\frac{1+\hat{v}^{3\alpha}\omega/\nu_{ei}}{i\omega\nu_{ei}+\frac{2}{3}s^{2}\hat{v}^{2+3\alpha}}e^{-\hat{v}^{2}}.\label{eq:cond_drakeLee}
\end{equation}
We now evaluate Eq. (\ref{eq:gladdmicro}) using Eq. (\ref{eq:cond_drakeLee}).
We first calculate the spatial integral, and then the velocity-space
integral, to obtain
\begin{equation}
\Delta^{\prime}\propto a_{1}\left(1-\frac{\omega_{*e}}{\omega}\right)-a_{2}\frac{\omega_{T}}{\omega}+i\frac{\omega}{\nu_{ei}}\left[a_{3}\left(1-\frac{\omega_{*e}}{\omega}\right)-a_{4}\frac{\omega_{T}}{\omega}\right],\label{eq:mah}
\end{equation}
where $a_{i}$ are real, positive constant functions of $\alpha$.
The fundamental frequency of the mode is then determined perturbatively
in $\delta/\rho_{i},$ as we did for Eq. (\ref{eq:coll_DT}) , solving
for 
\begin{equation}
a_{1}\left(1-\frac{\omega_{*e}}{\omega}\right)-a_{2}\frac{\omega_{T}}{\omega}=0.\label{eq:mahmah}
\end{equation}
When the collision frequency is energy-independent, $\alpha=0,$ we
obtain the familiar solution 
\begin{equation}
\omega=\omega_{*e}(1+\eta_{e}/2).
\end{equation}
In this case, $a_{4}=a_{2},$ and $a_{3}=a_{1},$ therefore the destabilising
collisional term on the RHS of Eq. (\ref{eq:mah}) cancels exactly
when evaluated at $\omega=\omega_{*e}(1+\eta_{e}/2).$ When $\alpha=1,$
the destabilising term remains, even when evaluated at the value of
$\omega$ which solves for Eq. (\ref{eq:mahmah}). We immediately
notice that the $\eta_{e}-$term on the RHS of Eq. (\ref{eq:mah})
does not cancel if the fundamental frequency of the mode \textsl{is
not }$\omega=\omega_{*e}(1+\eta_{e}/2).$ When $\omega>\omega_{*e}(1+\eta_{e}/2),$
the $\eta_{e}-$term is actually stabilising. However, when the constant-$A_{\parallel}$
approximation fails, $\omega\approx\omega_{*e}$ {[}see Eq. (\ref{eq:omCHZ}){]}
the destabilising term remains. We want to stress that, from gyrokinetic
simulations, we found that the energy-dependent collision frequency
can still play an important role in destabilizing an electron temperature
gradient driven mode. The simplest way to include this effect in our
theory is to modify the coefficient $\sigma_{0}$ in Eq. (\ref{eq:conduttivita})
by using a Pad\'e approximant
\begin{equation}
\sigma_{0}\rightarrow\sigma_{0}^{P}=1-\frac{\omega_{T}}{\omega}\left(1+\frac{\nu_{ei}/\omega}{1-i\nu_{ei}^{2}/\omega^{2}}\right)\,\,\,\mbox{for}\,\,\nu_{ei}\sim\omega,
\end{equation}
so that $\sigma_{0}\rightarrow1-\frac{\omega_{T}}{\omega}\,\,\,\mbox{for}\,\,\nu_{ei}\ll\omega,$
and $\sigma_{0}\rightarrow1-\frac{\omega_{T}}{\omega}\left(1+i\omega/\nu_{ei}\right)\,\,\,\mbox{for}\,\,\nu_{ei}\gg\omega.$

\subsection{Collisionless limits and ETG}

While we have been able to formulate the problem of microtearing modes
and relate it to the physics of the drift-tearing mode, the relationship
between these reconnecting modes and the collisionless electrostatic
ETG (which shares the same drive) still remains unclear. In order
to gain some insight into this aspect of the theory, we show firstly
that our Eqs. (\ref{eq:elcont})-(\ref{eq:grnohmslaw})-(\ref{eq:sum3})
support the electrostatic collisionless ETG in 3D shearless geometry.
Secondly, we consider the case of finite shear. We return to the eigenvalue
equations (\ref{eq:semi1}) and (\ref{eq:semi2}), and solve them
in a sound expansion, keeping $\hat{\beta}_{T}$ arbitrary, but considering
what we call the ``deeply-unstable ETG'' ordering, that is 
\[
\frac{k_{\parallel}^{2}v_{the}^{2}}{\omega^{2}}=s_{cl}^{2}\sim\frac{\omega}{\omega_{T}}\sim\varepsilon\ll1.
\]
The finite magnetic shear case gives a number of marginally stable
modes. We explain how this result relates to previous works \citep{coppiMT}.
We then perform a series of numerical simulations with the gyrokinetic
code {\tt GS2}, and find that the only electron temperature gradient
driven collisionless reconnecting mode is a tearing parity ETG, which
is mostly electrostatic.

\subsubsection{Collisionless ETG in KREHM: no shear\label{sub:Collisionless-ETG-in}}

We Fourier transform the $\nu_{ei}\rightarrow0$ limit of Eq. (\ref{eq:curr}),
and use Eq. (\ref{eq:land cond}), to obtain
\begin{equation}
\frac{k_{\perp}^{2}d_{e}^{2}}{2}=\left\{ \zeta^{2}-\frac{\tau}{Z}\frac{k_{\perp}^{2}d_{e}^{2}}{2}\right\} \left\{ 1+\zeta Z\left(\zeta\right)+\frac{1}{4}\frac{\omega_{T}}{\omega}\zeta\left[Z\left(\zeta\right)-2\zeta\left(1+\zeta Z\left(\zeta\right)\right)\right]\right\} ,\label{eq:collessLan}
\end{equation}
where $\zeta=\omega/(k_{z}v_{the}).$ When $\hat{\omega}=\omega_{T}/\omega\equiv0,$
this is Eq. (B.12) of Ref. \citep{zocco:102309}. In this case, for
$\zeta\ll1,$ Eq. (\ref{eq:collessLan}) gives a damped kinetic Alfv\'en
wave \citep{zocco:102309}. For $k_{\perp}^{2}d_{e}^{2}\gg\zeta^{2}=\omega^{2}/(k_{z}^{2}v_{the}^{2})\sim\omega_{T}/\omega\gg1,$
we obtain$^{1}$ \footnotetext[1]{The electrostatic limit of our
model is found for $k_{\perp}d_{e}\gg1.$ Indeed, by using Ampere's
law and the electron continuity equation, we find that
\begin{equation}
k_{\perp}^{2}d_{e}^{2}\frac{e}{m_{e}c}A_{\parallel}=u_{\text{\ensuremath{\parallel},e }}\sim\frac{\omega}{k_{\parallel}}\frac{e\varphi}{T_{0e}}.
\end{equation}
When the parallel electron dynamics is included, $\omega\sim k_{\parallel}v_{the},$
we obtain
\begin{equation}
\frac{v_{the}}{c}A_{\parallel}\sim\frac{1}{k_{\perp}^{2}d_{e}^{2}}\varphi,
\end{equation}
thus, the electromagnetic component of the electron gyrokinetic potential,
$\chi=\varphi-(v_{\parallel}/c)A_{\parallel},$ is negligible in the
$k_{\perp}^{2}d_{e}^{2}\gg1$ limit.}
\begin{equation}
1\approx-\frac{\tau}{Z}\frac{1}{4}\frac{k_{z}^{2}v_{the}^{2}\omega_{T}}{\omega^{3}}\left\{ 1-2\zeta^{5}i\sqrt{\pi}\sigma e^{-\zeta^{2}}\right\} ,
\end{equation}
where $\sigma=0$ for $\Im[\zeta]>\left|\Re[\zeta]\right|^{-1},$
$\sigma=1$ for $\Im[\zeta]<\left|\Re[\zeta]\right|^{-1}$ $,$ and
$\sigma=2$ for $\Im[\zeta]<-\left|\Re[\zeta]\right|^{-1}.$ The asymptotic
expansion for large $\zeta$ is in the complex plane, and a direction
for the limit has to be specified. If we choose $\arg\zeta=\pi/3,$
there is always a critical $\zeta_{c}$ above which $\Im[\zeta]>\left|\Re[\zeta]\right|^{-1}$
and $\sigma=0.$ Therefore, for 
\begin{equation}
\frac{\tau}{Z}\frac{1}{4}\frac{\omega_{T}^{2}}{k_{z}^{2}v_{the}^{2}}>\left(\frac{4\sqrt{3}}{3}\right)^{3/2},
\end{equation}
 we find the electrostatic collisionless ETG mode 
\begin{equation}
\omega_{0}^{3}\approx-\frac{\tau}{Z}\frac{1}{4}k_{z}^{2}v_{the}^{2}\omega_{T}.
\end{equation}

\subsubsection{Electron sound expansion: finite shear\label{sub:Electron-sound-expansion:}}

In the case of finite shear and $k_{z}\equiv0$, the analysis is more
complex. We write Eq. (\ref{eq:curr}) for the magnetic potential
(from now on $s_{cl}\equiv s$) 
\begin{equation}
\frac{F_{\infty}-s^{2}\hat{\sigma}_{L}}{F_{\infty}\hat{\sigma}_{L}}\left(\frac{d^{2}A_{\parallel}}{ds^{2}}-k_{y}^{2}\delta^{2}A_{\parallel}\right)=-\hat{\beta}_{T}\hat{\omega}^{2}A_{\parallel}.\label{eq:machecazzoneso-1}
\end{equation}
 In the limit $k_{\parallel}^{2}v_{the}^{2}/\omega^{2}=s^{2}\sim\frac{\omega}{\omega_{T}}\sim\varepsilon\ll1,$
at scales $s\sim(\omega/\omega_{T})^{1/2}\ll1,$ we obtain 
\begin{equation}
\frac{d^{2}A_{\parallel}}{d\xi^{2}}=\delta_{MT}^{2}k_{y}^{2}A_{\parallel}-\frac{A_{\parallel}}{1+\alpha^{2}\xi^{2}},\label{eq:mia}
\end{equation}
 where $\alpha^{2}=\frac{\tau}{Z}/\left(\hat{\beta_{T}}\hat{\omega}^{2}\right),$
and 
\begin{equation}
\xi^{2}=\frac{1}{2}\hat{\omega}\hat{\beta}_{T}s^{2}\rightarrow\xi=\sqrt{\frac{\hat{\beta}_{T}}{2}}\frac{k_{y}v_{the}}{\sqrt{\omega\omega_{T}}}\frac{x}{L_{s}}.\label{eq:MTscale}
\end{equation}
 Equation (\ref{eq:MTscale}) shows us that there is a new scale,
\begin{equation}
\delta_{E=}\sqrt{\frac{2}{\hat{\beta}_{T}}}\frac{\sqrt{\omega\omega_{T}}}{k_{y}v_{the}}L_{s},
\end{equation}
 that determines the width of the mode. The electron sound expansion
is valid for 
\begin{equation}
s\sim\sqrt{\omega/\omega_{T}}\ll1.
\end{equation}
 When $\omega\approx\omega_{T},$ the new scale $\delta_{E}$ coincides
with the electron inertial scale $d_{e},$ and Eq. (\ref{eq:mia})
is no longer valid. Equation (\ref{eq:mia}) corresponds to Eq. (3)
of Ref. \citep{coppiMT}. It has been used to prove the existence
of a collisionless microtearing mode when finite $k_{y}d_{e}$ is
considered in Ohm's law \citep{coppiMT} . We now solve it exactly
in the two asymptotic limits $k_{y}^{2}\delta_{E}^{2}\ll1,$ and $k_{y}^{2}\delta_{E}^{2}\sim1.$

\paragraph{Small $\delta_{E}^{2}k_{y}^{2}$ solution}

The even parity solution of Eq. (\ref{eq:mia}), for $\delta_{E}^{2}k_{y}^{2}\ll1,$
is 
\begin{equation}
A_{\parallel}\left(\xi^{2}\right)=_{2}F_{1}\left(-\frac{1}{4}-\mu,\,-\frac{1}{4}+\mu;\,\frac{1}{2}\,;-\alpha^{2}\xi^{2}\right),\label{eq:solmia}
\end{equation}
 where $_{2}F_{1}$ is the hypergeometric function, and $\mu=\sqrt{\alpha^{2}-4}/(4\alpha).$
We look for an eigenvalue of the form $\hat{\omega}=\hat{\omega}_{0}+i\hat{\gamma},$
with $\hat{\gamma}\ll\hat{\omega}_{0},$ hence $\mu<1/4.$ This forces
$\hat{\beta}_{T}$ to be small.

For $\alpha^{2}\xi^{2}\gg1,$ when $1/(\alpha\xi)\sim\delta_{E}k_{y}\ll1,$
the finite wavenumber of the perturbation can be important, and Eq.
(\ref{eq:mia}) becomes 
\begin{equation}
\frac{d^{2}A_{\parallel}}{d\xi^{2}}=\left(\delta_{E}^{2}k_{y}^{2}-\frac{1}{\alpha^{2}\xi^{2}}\right)A_{\parallel}.\label{eq:jim}
\end{equation}
 The solution of Eq. (\ref{eq:jim}) that decays exponentially at
infinity is 
\[
A_{\parallel}=C\sqrt{\xi}K_{2\mu}\left(\delta_{E}k_{y}\xi\right),
\]
 where $C$ is a constant. After matching to the solution (\ref{eq:solmia}),
we obtain the following dispersion relation 
\begin{equation}
\frac{\Gamma^{2}\left(2\mu\right)\Gamma^{2}\left(-\frac{1}{4}-\mu\right)}{\Gamma^{2}\left(-2\mu\right)\Gamma^{2}\left(-\frac{1}{4}+\mu\right)}\frac{\frac{1}{4}+\mu}{\frac{1}{4}-\mu}=\left(\frac{\omega^{3}}{2\tau v_{the}^{2}/L_{s}^{2}\omega_{T}}\right)^{2\mu}.\label{eq:fullmu}
\end{equation}

\paragraph*{Relation with the low$-\hat{\beta}_{T}$ theory}

In the $\hat{\beta}_{T}\ll1$ limit, Eq. (\ref{eq:fullmu}) reduces
to 
\begin{equation}
\left(\mu-\frac{1}{4}\right)\pi\sqrt{\frac{1}{2}\frac{\tau}{Z}\frac{1}{\hat{\omega}}}=-\hat{\omega}\delta_{*}k_{y},\label{eq:kaibuzzu}
\end{equation}
where we recall $\delta_{*}=\omega_{T}/(k_{y}v_{the})L_{s}$. It is
easy to verify that, for $\Delta^{\prime}=-2k_{y},$ Eq. (\ref{eq:kaibuzzu})
is exactly 
\begin{equation}
\left(\frac{\Delta^{\prime}\delta_{*}}{\pi\hat{\beta}}\right)^{2}=\frac{1}{2}\hat{\omega}\frac{Z}{\tau}.\label{eq:coppolo-1-1}
\end{equation}
 We must derive this result from the $\omega/\omega_{T}\ll1$ limit
of (\ref{eq:disprelfinal}), which was derived in a low$-\hat{\beta}_{T}$
expansion. When ions are unmagnetised, and the ideal MHD drive is
not too large (constant$-A_{\parallel}$ approximation), $C\approx1,$
we have 
\begin{equation}
\sqrt{2\frac{\nu_{ei}}{\omega_{T}}}e^{-i\frac{\pi}{4}}\frac{\delta_{*}\Delta^{\prime}}{\pi\hat{\beta}_{T}}{\hat{\omega}}^{1/2}=-\frac{2}{\pi}{\hat{\omega}}^{2}\int_{0}^{\infty}ds\frac{F_{\infty}\hat{\sigma}_{e}}{F_{\infty}\left(1-i\frac{\hat{\omega}}{\hat{\nu_{ei}}}\right)-s^{2}\hat{\sigma}_{e}}.\label{eq:mado}
\end{equation}
 Computing the integral on the RHS is particularly easy using the
sound expansion limit \citep{coppiMT} $k_{\parallel}v_{the}/\omega\equiv s\ll1$.
Let us write, 
\begin{equation}
I_{e}=-\frac{Z}{\tau}\int_{0}^{\infty}ds\frac{1-\hat{\omega}^{-1}}{\left(1-i\frac{\hat{\omega}}{\hat{\nu}_{ei}}\right)\frac{Z}{\tau}+s^{2}\left(1-\hat{\omega}^{-1}\right)}.\label{eq:echecazzo}
\end{equation}
 Within this expansion in $s\ll1,$ we consider $Z/\tau\ll1$ in order
to keep the quadratic term $s^{2}(1-\hat{\omega}^{-1})\sim Z/\tau\ll1$
in the denominator of the integrand, which continues to remain convergent
in all our subsidiary expansions. Essentially, the electrostatic potential
is \textit{never} neglected and the current is \textit{always} proportional
to the electric field in the electron region.

Then, we obtain 
\[
I_{e}=-\frac{\pi}{2}\sqrt{\frac{Z}{\tau}\left(1-\frac{1}{\hat{\omega}}\right)/\left(1-i\omega/\nu_{ei}\right)},
\]
 and we obtain the following dispersion relation 
\begin{equation}
\sqrt{2\frac{\nu_{ei}}{\omega_{T}}}e^{-i\frac{\pi}{4}}\frac{\delta_{*}\Delta^{\prime}}{\pi\hat{\beta}_{T}}\hat{\omega}^{1/2}=-{\hat{\omega}}^{2}\sqrt{\frac{Z}{\tau}\frac{1-\hat{\omega}^{-1}}{1-i\frac{\omega}{\nu_{ei}}}}.
\end{equation}
 In the limit $\frac{1}{\sqrt{N}}\sim\frac{Z}{\tau}\ll\frac{\nu_{ei}}{\omega}\ll1,$
the collision frequency exactly cancels and we obtain 
\begin{equation}
-\left(\frac{\Delta^{\prime}\delta_{*}}{\pi\hat{\beta}}\right)^{2}=\frac{1}{2}\hat{\omega}^{2}\frac{Z}{\tau}\left(1-\frac{1}{\hat{\omega}}\right).\label{eq:coppolo}
\end{equation}
 One can get the same result by using the truly collisionless equation
(\ref{eq:machecazzoneso-1}), and noticing that $\hat{\sigma}_{L}(0)=-1/2\hat{\sigma}_{e}(0).$

Taking the $\omega/\omega_{T}\ll1$ limit, we obtain Eq. (\ref{eq:coppolo-1-1}).
We have thus proved that the limits $\hat{\omega}\ll1$ and $\hat{\beta}_{T}\ll1$
commute, and taking \textit{both} limits results in a stable wave
oscillating with the frequency 
\begin{equation}
\hat{\omega}_{0}=\left(\frac{2}{\pi}\right)^{2}2\frac{\tau}{Z}\left(\frac{k_{y}\delta_{*}}{\hat{\beta}_{T}}\right)^{2}.\label{eq:freqmia}
\end{equation}
We conclude that there is no collisionless microtearing mode in this
low $\hat{\beta}_{T}$ limit.

\paragraph{Finite $k_{y}^{2}\delta_{E}^{2}\sim1$ limit}

Let us prove that a localised perturbation of the type $\exp[-\sigma x^{2}]$
fails to give reconnection, even if we retain the electrostatic potential
and we consider $k_{y}^{2}\delta_{E}^{2}\sim1.$

For $\alpha^{2}\xi^{2}\ll1,$ Eq. (\ref{eq:mia}) reduces to
\begin{equation}
\frac{d^{2}A_{\parallel}}{dY^{2}}=\frac{1}{\alpha^{2}}\left\{ \hat{b}-1+Y^{2}\right\} A_{\parallel},\,\,\,\mbox{for}\, Y\ll1,\label{eq:echeminchia}
\end{equation}
which implies $Y^{2}\equiv\alpha^{2}\xi^{2}\sim1-\hat{b}\ll1.$

If we use the ansatz
\begin{equation}
A_{\parallel}=e^{-\lambda Y^{2}},
\end{equation}
with $\Re(\lambda)>0,$ we obtain
\begin{equation}
\lambda=+\frac{1}{2\alpha}=+\frac{1}{2}\sqrt{\frac{\hat{\beta}_{T}}{\tau/Z}}\hat{\omega},
\end{equation}
for an electron mode, and
\begin{equation}
\hat{b}-1=-\sqrt{\frac{\tau/Z}{\hat{\beta}_{T}}}\frac{1}{\hat{\omega}}.
\end{equation}
This is equivalent to
\begin{equation}
\left(\hat{\omega}k_{y}^{2}d_{e}^{2}-1\right)\hat{\omega}\sqrt{\frac{\hat{\beta}_{T}}{\tau/Z}}=-1.\label{eq:neweq}
\end{equation}
There are two maginally stable solutions consistent with the conditions
$\hat{b}\approx1,$ and $\hat{\omega}\ll1:$
\begin{equation}
\hat{\omega}_{1}\approx\frac{1}{k_{y}^{2}d_{e}^{2}},
\end{equation}
and
\begin{equation}
\hat{\omega}_{2}\approx2\sqrt{\frac{\tau}{\hat{\beta}_{T}}}.
\end{equation}
In this case, the eigenfunction decays as $\exp[-\alpha x^{2}/(2\delta_{E}^{2})]$
for $x\gg\delta_{E}/\alpha.$ 

In principle, we could impose the tearing-stable MHD boundary condition
at large $x,$ if we match the low$-k$ solution of Eq. (\ref{eq:mia})
to the Fourier transform of $A_{\parallel}^{MHD}=\exp[-k_{y}\left|x\right|].$
Let us Fourier transform Eq. (\ref{eq:echeminchia}), to obtain
\begin{equation}
\frac{d^{2}A_{\parallel}(\theta)}{d\theta^{2}}=\left\{ \left(\hat{b}-1\right)+\alpha^{2}\theta^{2}\right\} A_{\parallel}(\theta),
\end{equation}
 where $\theta=\alpha^{-1}\delta_{MT}k$ is the Fourier conjugate
of $Y.$ The solution of this equation which is even for all $\rho=\frac{1}{4}+\frac{\hat{b}-1}{4\alpha}$
is
\begin{equation}
A_{\parallel}(\theta)=A_{0}e^{-\frac{1}{2}\frac{\theta^{2}}{\alpha}}\,_{1}F_{1}\left(\rho;\frac{1}{2};\frac{1}{\alpha}\theta^{2}\right).
\end{equation}
On the other hand, the Fourier transform of the tearing-stable boundary
condition is
\begin{equation}
A_{\parallel}=\frac{A_{0}}{k_{y}^{2}+k^{2}}\approx\frac{A_{0}}{k_{y}^{2}}\left(1-\frac{k^{2}}{k_{y}^{2}}\right).
\end{equation}
 Thus, after matching these two expressions in the low$-\theta$ limit,
we find
\begin{equation}
\hat{b}\left(\hat{b}-1\right)=-2\alpha^{2}.
\end{equation}
Explicitly, we have
\begin{equation}
\sqrt{\frac{2}{\hat{\beta}_{T}}}\frac{\omega}{v_{the}/L_{s}}k_{y}d_{e}-1=-\tau\sqrt{\frac{\hat{\beta}_{T}}{2}}\frac{k_{y}d_{e}}{\omega^{3}/(v_{the}/L_{s})^{3}}.
\end{equation}
One of the roots that solves for this equation is connected to the
unstable root
\begin{equation}
\frac{\omega^{4}}{(v_{the}/L_{s})^{4}}\approx-\tau\frac{\hat{\beta}_{T}}{2},
\end{equation}
for $\hat{b}\gg1.$ However, we cannot accept this root within our
analysis, as the derivation is only valid for $\hat{b}\approx1.$

\section{Gyrokinetic simulations in the collisionless limit\label{sec:Numerical-benchmark-for}}

In Section (\ref{sec:Microtearing-Mode}) we proved that, for a tearing-stable
configuration and finite collisions, there exists an unstable microtearing
mode. This mode becomes stable in the collisionless limit. In the
following, we show numerically that, even if the microtearing mode
is stabilised when collisions are small, a tearing-parity, mostly
electrostatic ETG, can still cause magnetic reconnection. 

We performed a series of gyrokinetic simulations using the code {\tt GS2}
We show the cases in which $\hat{\beta}_{T}=0.1,$ and $\nu_{ei}/(v_{the}/L_{s})=0.09.$
In slab {\tt GS2}, the quantity $L_{s}$ defines the variable $k_{p}=L_{ref}/(\hat{s}L_{s}),$
which implies that our box has parallel length $L_{z}=\hat{s}L_{s}/(2\pi).$
Here $\hat{s}$ is the local magnetic shear, and $L_{ref}$ a reference
legth. As the parallel scale of the mode depends on $k_{y},$ we need
to change $L_{z}$ as we change $k_{y}$. To do this we change $\hat{s}$.
At the same time we change $k_{p}$ to keep $L_{s}$ constant. Then,
for $\beta_{e}=m_{e}/m_{i}=1/2345,$ in terms of input parameters,
we have $\hat{\beta}_{T}=(m_{e}/2m_{i})(L_{ref}/L_{T})^{2}(1/(k_{p}\hat{s}))^{2},$
with $L_{ref}/L_{T}=50,$ and $k_{p}\hat{s}=1.85.$

\begin{figure}[!h]
\input{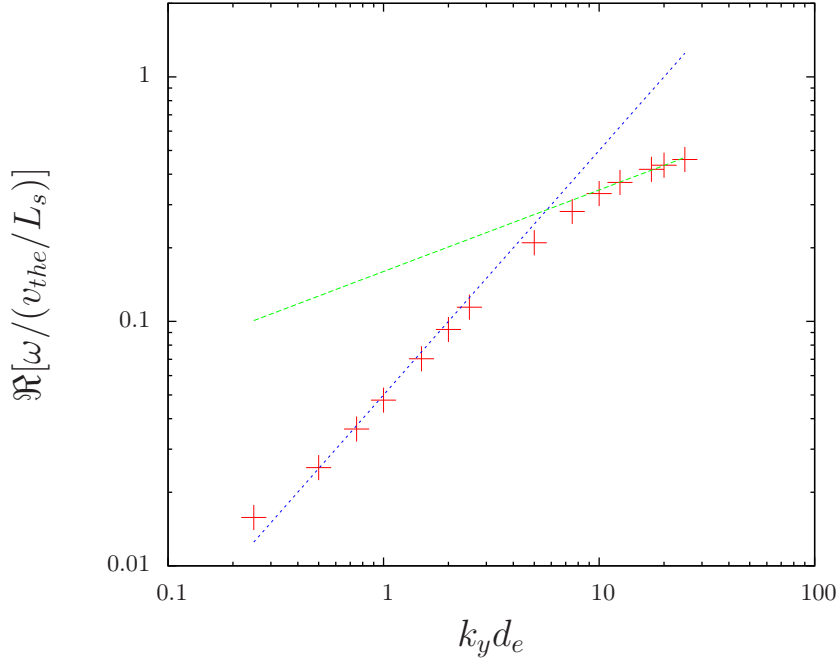} 

\caption{The spectrum of the real part of the eigenvalue obtained from gyrokinetic
electromagnetic simulations using the code {\tt GS2}. Here $\hat{\beta}_{T}=0.1$
and $\nu_{ei}/(v_{the}/L_{s})=0.09.$ The mode is essentially collisionless.
The two lines are the curves $\omega=0.05\,\omega_{T}$ and $\omega/(v_{the}/L_{s})=0.16\left[\omega_{T}/(v_{the}/L_{s})\right]^{1/3}.$}
\label{fig:realGS2}
\end{figure}

\begin{figure}[!h]
\input{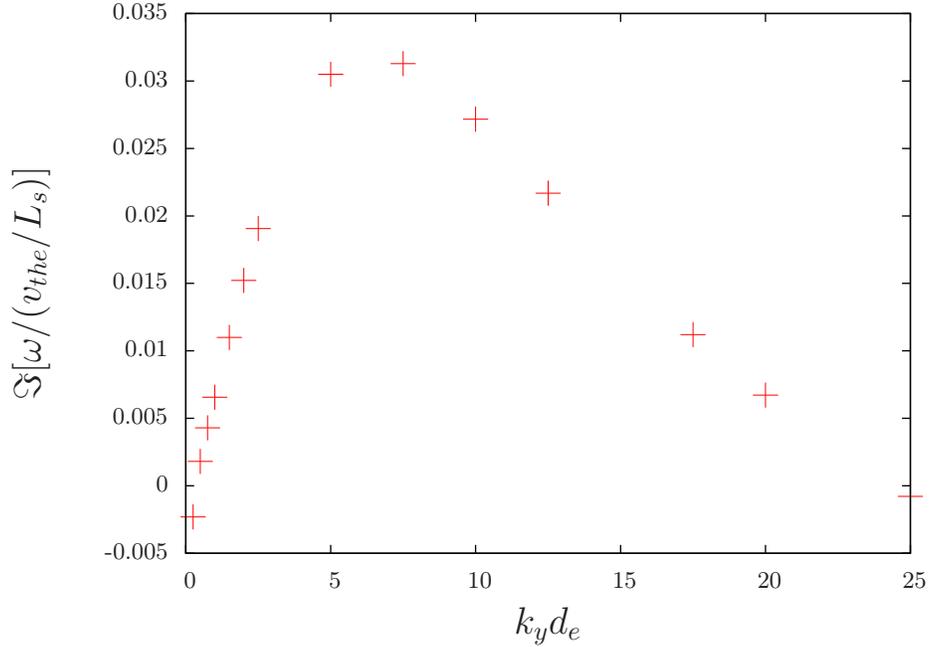} 

\caption{The spectrum of the imaginary part of the eigenvalue obtained from
gyrokinetic electromagnetic simulations using the code {\tt GS2}.
Here $\hat{\beta}_{T}=0.1,$ and $\nu_{ei}/(v_{the}/L_{s})=0.09.$
The mode is essentially collisionless. }
\label{fig:imGS2}
\end{figure}

\begin{figure}[!h]
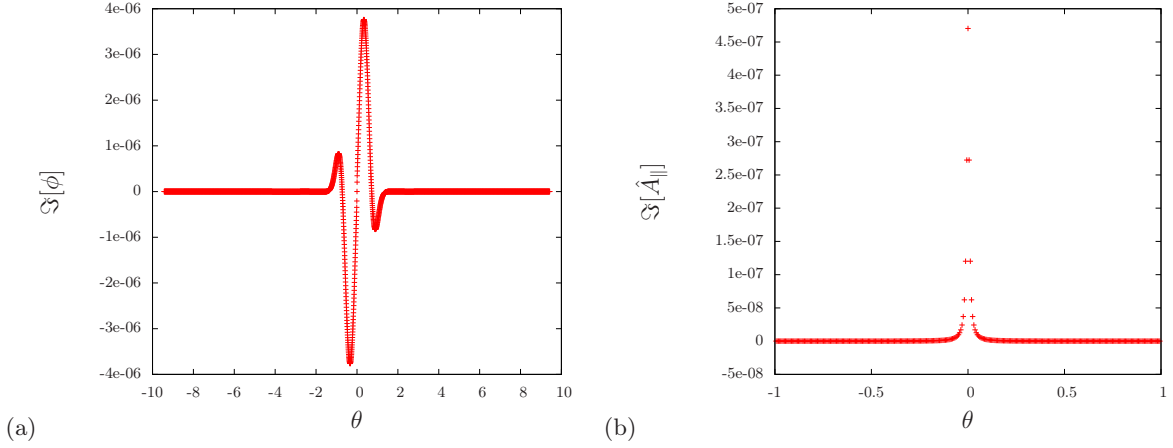

\subfloat{(a)}{\scalebox{0.65}{\input{eigenDT.tex}}} \subfloat{(b)}{\scalebox{0.65}{\input{eigenDTApar.tex}}} 

\caption{Eigenfunctions in the drift tearing regime: $k_{y}d_{e}=0.5\times\sqrt{2}.$
Imaginary part of $\text{\ensuremath{\phi}}$ (a), and $\hat{A}_{\text{\ensuremath{\parallel}}}$
(b). }
\label{fig:eigenDT}
\end{figure}

\begin{figure}[!h]
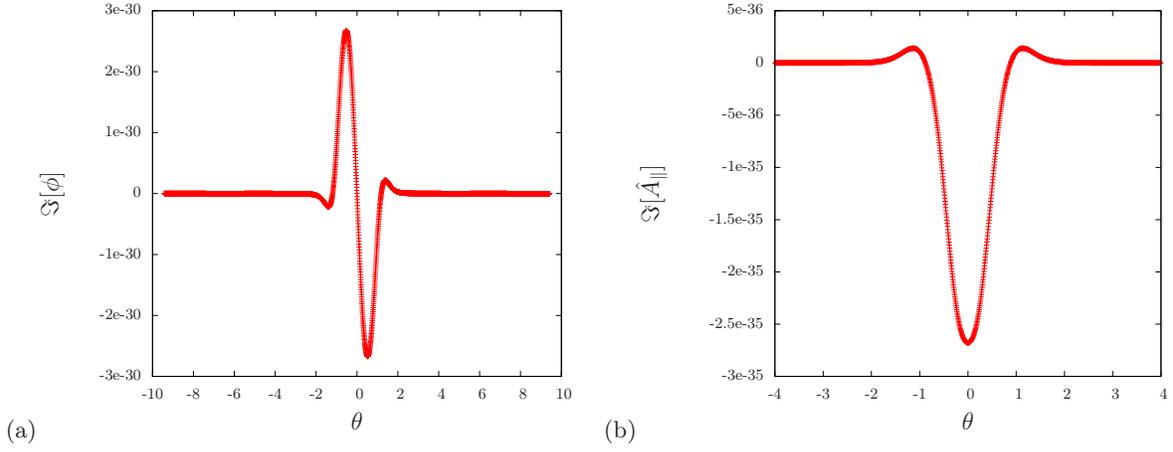

\subfloat{(a)}{\scalebox{0.65}{\input{eigenETG.tex}}} \subfloat{(b)}{\scalebox{0.65}{\input{eigenETGApar.tex}}} 

\caption{Eigenfunctions in the ETG regime: $k_{y}d_{e}=25.\times\sqrt{2}.$
Imaginary part of $\text{\ensuremath{\phi}}$ (a), and $\hat{A}_{\text{\ensuremath{\parallel}}}$
(b). }
\label{fig:eigenETG}
\end{figure}

\begin{figure}[!h]
\input{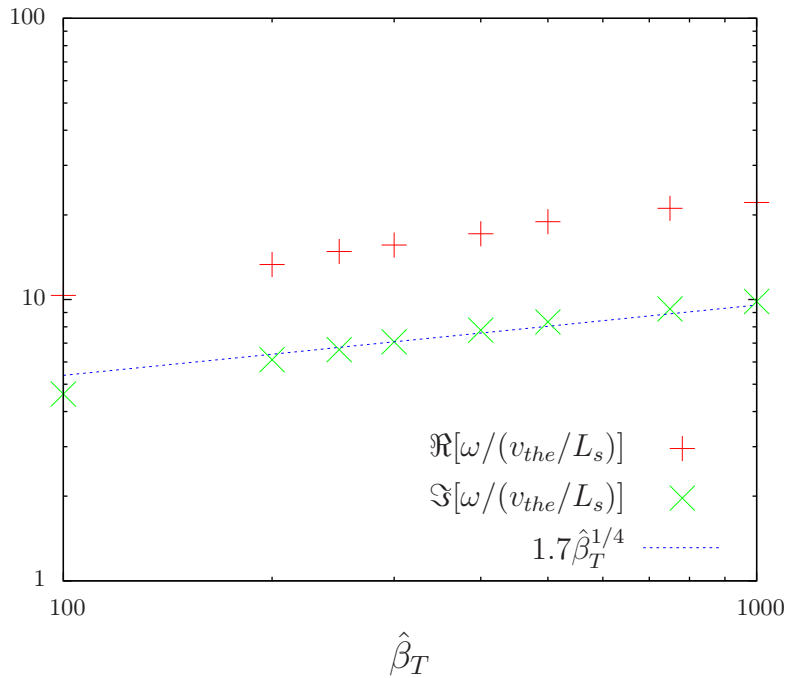} 

\caption{The scaling with $\hat{\beta}_{T}$ for the eigenvalue in Figs. (\ref{fig:realGS2})
and (\ref{fig:imGS2}) in the regime $\omega\sim\omega_{T},$ $k_{y}d_{e}=0.1\times\sqrt{2}.$ }
\label{fig:betascan}
\end{figure}

For this collisionality, the mode is basically collisionless. For
these parameters, we observe that the response of the ions is nearly
adiabatic. We stress that we choose $\beta_{e}=m_{e}/m_{i}=1/2345$
in order to enforce the ordering used to derive Eqs. (\ref{eq:elcont})-(\ref{eq:grnohmslaw})-(\ref{eq:sum3}).
A finite amount of collisions is required to ensure regular and convergent
eigenfunctions. The eigenfunctions are shown in Figs. (\ref{fig:eigenDT})-(\ref{fig:eigenETG}).
Here $\phi=(L_{ref}/\rho_{ref})Z_{ref}e\varphi/T_{ref},$ and $\hat{A}_{\parallel}=v_{the}(L_{ref}/\rho_{ref})Z_{ref}eA_{\parallel}/T_{ref},$
with $T_{ref}=T_{e},$ $m_{ref}=m_{i},$ and $\theta$ is the conventional
angle-like variable of ballooning theory \citep{PhysRevLett.40.396}.
The velocity space resolution is $8$ grid points in energy $E$,
and $16$ in the pitch-angle variable $\lambda=\mu/E,$ where $\mu=mv_{\perp}^{2}/(2B)$
is the magnetic moment. The wave-numbers resolved are in the range
$0.1\leq k_{y}d_{e}\leq30.$ The spectra are in Figs. (\ref{fig:realGS2})-(\ref{fig:imGS2}).
After inspection of the real frequency spectrum, we identify two different
regimes: a drift-tearing regime $\omega\sim\omega_{T},$ for $k_{y}d_{e}\lesssim5,$
and an ETG regime $\omega\sim\left[(v_{the}/L_{s})^{2}\omega_{T}\right]^{1/3},$
for $k_{y}d_{e}\gtrsim10,$ see Fig. (\ref{fig:realGS2}). 
In both regimes, it is easy to verify that an electrostatic calculation
would give qualitatively the same results as Figs. (\ref{fig:realGS2})
and (\ref{fig:imGS2}). We conclude that the unstable mode presented
in Figs. (\ref{fig:imGS2}) and (\ref{fig:realGS2}) is a mostly electrostatic
tearing parity ETG that generates magnetic reconnection at the electron
scale. We notice, however, that the ratio of the amplitudes of the
electromagnetic and electrostatic component varies greatly in the
different regimes identified here. Indeed, $\hat{A}_{\parallel}/\phi$
ranges from $10^{-1}$ in the drift-tearing to $10^{-5}$ in the ETG
regime. The mode is stabilised by collisions (not shown). An analysis
similar to that of Connor et al. \citep{0741-3335-55-12-125003} could
predict how much reconnection this mode is actually generating. We
leave this question open for the moment, and numerically derive a
scaling with $\hat{\beta}_{T}$ for the growth rate in the $\omega\sim\omega_{T}$
regime. We then choose one simulation from Fig (\ref{fig:realGS2})
and, for a given $k_{y}d_{e}=0.1\times\sqrt{2},$ we perform a scan
in $\hat{\beta}_{T},$ keeping $\beta_{e}=m_{e}/m_{i}$ constant.
The results are shown in Fig. (\ref{fig:betascan}). The scaling $\gamma\sim\omega\sim\hat{\beta}_{T}^{1/4}$
is observed, even if only for unrealistically large values of $\hat{\beta}_{T}.$
To overcome nomenclature issues, we resist the temptation to call
the drift-tearing branch a collisionless microtearing mode.

\section{Conclusions and Discussion}

In our efforts to create a new, simple description of fine-scale,
kinetic electromagnetic turbulence in plasma, in Sec. (\ref{sec:Model-equations})
we derived a hybrid fluid-kinetic model, for micro and macro reconnecting
modes and electron turbulence, which is based on gyrokinetic theory
\citep{zocco:102309}. The model features a gyrokinetic Poisson law
for electrostatic perturbations {[}Eq. (\ref{eq:GKPL}) used in Eq.
(\ref{eq:grnohmslaw}){]}, and the electron parallel momentum equation
{[}Eq. (\ref{eq:grnohmslaw}){]} in the form of a generalised Ohm's
law. This is coupled to electron kinetics via electron temperature
fluctuations defined by Eqs. (\ref{eq:delT}) and (\ref{eq:sum3}). 

The derivation of the linearised equation for a sheared magnetic equilibrium
{[}Eqs. (\ref{eq:lin1}) and (\ref{eq:lin3}){]} is carried out in
Sec. (\ref{sub:Linear-eigenvalue-problem}). In Sec. (\ref{sub:Electron-conductivity})
we presented a new solution for the electron kinetic problem {[}Eq.
(\ref{eq:gensolkin}){]}, based on a Hermite expansion of the electron
distribution function {[}Eq. (\ref{eq:exp}){]}. This generated a
general electron conductivity {[}Eq (\ref{eq:conduttivita}){]} which
reproduces the well-known highly collisional limit for non-isothermal
(semicollisional) electrons {[}Eq. (\ref{eq:cond}){]}, and allows
us to represent reasonably well the exact collisionless results with
a finite number of Hermite moments {[}Figs. (\ref{fig:sigmaM6})-(\ref{fig:sigmaM20}){]}.
This, in turn, made it possible to include a small but finite amount
of collisions in the study of kinetic reconnecting modes, and opened
the way to our description of kinetic microtearing modes {[}Sec (\ref{sec:Microtearing-Mode}){]}.
The theory of these is largely based on the theory of tearing modes
{[}Sec. (\ref{sec:Dispersion-Relation}){]} that features the new
electron conductivity, and retains gyrokinetic ions {[}Eq. (\ref{eq:gendisprel}){]}.
In the weakly collisional regime, for tearing-unstable configurations,
we benchnmarked analytical and semianalytical results against a hybrid
fluid-kinetic code ({\tt Viriato}) and the gyrokinetic code {\tt AstroGK}
{[}Fig. (\ref{fig:virAGKbench-1}){]}. For tearing-stable configurations,
our analysis uncovers a series of interesting facts. The inclusion
of the electrostatic potential introduces a new fundamental parameter
in the theory, which is the ratio of the kinetic electron and ion
scales $d_{e}/\rho_{s}.$ Within our formulation, it is particularly
easy to see the fundamental role of electron inertia in enabling magnetic
reconnection when the mode is unstable {[}Fig. (\ref{fig:node}){]}.
We also see, however, that a finite collisionality is always needed
to produce an unstable kinetic microtearing mode, and the mode seems
to be marginally stable when the general conductivity is close enough
to the exact result at zero collisionality {[}Eq. (\ref{eq:land cond}){]}.
Our analysis does not require an energy-dependent collision frequency,
but does not exclude its importance in driving some electron temperature
gradient driven modes. We identified the importance of the breaking
of the constant-$A_{\parallel}$ approximation in driving the microtearing
mode unstable, derived a high-$\hat{\beta}_{T}$ theory and gave analytical
expressions for the eigenmode in Eqs. (\ref{MTcoll}), (\ref{eq:MTcollless}),
and (\ref{eq:omCHZ})-(\ref{eq:gammaCHZ}). The critical electron
temperature gradient for instability is given in Eq. (\ref{eq:crieta}).

When the collisionality becomes smaller, one is justified in asking
whether any other micro mode will cause magnetic reconnection, since
this kind of microtearing mode does not seem to survive. The answer
is yes. In Sec. (\ref{sub:Collisionless-ETG-in}) we show, using the
gyrokinetic code {\tt GS2}, that the only electron temperature driven
collisionless reconnecting mode we could identify is a tearing parity
electromagnetic ETG which happens to reconnect, and is mostly electrostatic.
We identified two different branches of this mode: a drift-tearing
branch, for which $\omega\sim\omega_{T},$ and an ETG branch, for
which $\omega\sim\omega_{T}^{1/3}(v_{the}/L_{s})^{2/3}.$ In our analysis,
we neglected toroidal effects. While some authors found a toroidal
electron temperature gradient driven collisionless reconnecting mode
\citep{1.4799980}, its connection to the toroidal branch of the electrostatic
ETG remains unclear. Other short wavelength temperature gradient driven
instabilities are known to exist \citep{PhysRevLett.89.125005,smol1,smol2},
but their relation to our theory is not yet understood. Understanding
these aspects, with or without geometry, would require a theory of
ETG-driven magnetic reconnection {[}as was done in Ref. \citep{0741-3335-55-12-125003}
for the ion temperature gradient driven mode{]}. Only with such a
theory, we will be able to quantify how much electromagnetic transport
is caused by micro-reconnecting modes, and to investigate the relation
between the drift-tearing branch found in our gyrokinetic numerical
analyis {[}Fig. (\ref{fig:realGS2}){]}, and the solution of the new
dispersion relations Eqs. (\ref{eq:disprelfinal}) and (\ref{eq:disphighb}).
The numerical solution of Eqs. (\ref{eq:elcont})-(\ref{eq:sum3})
will definitely help us to answer these and other questions. 
\begin{acknowledgments}
We thank AA Schekochihin, Ken McClements, and F Porcelli for insightful
comments, JW Connor RJ Hastie for their contribution to the collisionless
theory, FI Parra in particular for a discussion on the marginal stability
of the fundamental root of the collisionless drift-tearing mode, JWC,
RJH, Martin Valovi\v c and Per Helander for their support throughout
the project. This work has received funding from the European Union's
Horizon 2020 research and innovation programme under grant agreement
number 633053 and from the RCUK Energy Programme {[}grant number EP/I501045{]}.
NFL has been supported by the Fundação para a Ciência e Tecnologia
grants Pest-OE/SADG/LA0010/2011, IF/00530/2013 and PTDC/FIS/118187/2010.
RN has been supported by JSPS KAKENHI Grant Number 24740373. The views
and opinions expressed herein do not necessarily reflect those of
the European Commission.
\end{acknowledgments}
\appendix

\section{Derivation of the model equations}

We derive the equations of the model for a somewhat general case,
that is for $\omega_{*e}=\frac{1}{2}k_{y}v_{the}\rho_{e}/L_{n}\neq0.$
This requires a considerable amount of algebra to treat the nonlinear
ion kinetic equation; however we found no other way to benchmark our
model against previous linear results \citep{coppi-collisionless,antonsen-coppi,pegoraro:478,0741-3335-54-3-035003}.

\subsection{Electrons}

We start with the electron kinetic equation \citep{frieman:502} for
$h_{e}=F_{e}-F_{0e}(1+e\varphi/T_{0e})$

\begin{equation}
\begin{split} & \frac{\partial h_{e}}{\partial t}+\mathbf{v}_{E}\cdot\nabla h_{e}+v_{\parallel}\hat{\mathbf{b}}\cdot\nabla h_{e}=-\frac{eF_{0e}}{T_{0e}}\frac{\partial}{\partial t}\left(\varphi-\frac{v_{\parallel}A_{\parallel}}{c}\right)\\
 & -\frac{c}{B_{0}}\mathbf{e}_{z}\cdot\nabla\left(\varphi-\frac{v_{\parallel}A_{\parallel}}{c}\right)\times\nabla F_{0e}+\left(\frac{\partial h_{e}}{\partial t}\right)_{coll},
\end{split}
\label{eq:edk}
\end{equation}
 where $\mathbf{v}_{E}=cB_{0}^{-1}(-\partial_{y}\varphi\mathbf{e}_{x}+\partial_{x}\varphi\mathbf{e}_{y})$
is the $\mathbf{E}\times\mathbf{B}$ drift velocity, $\hat{\mathbf{b}}\cdot\nabla=\partial_{z}-B_{0}^{\text{-1}}\{A_{\parallel},\},$
and 
\begin{equation}
F_{0e}=\frac{n_{0e}(\mathbf{x})}{\left[\pi v_{the}^{2}(\mathbf{x})\right]^{3/2}}e^{-\frac{v_{\parallel}^{2}+v_{\perp}^{2}}{v_{the}^{2}(\mathbf{x})}}
\end{equation}
is the inhomogeneous Maxwellian equilibrium, with temperature $T_{0e}(\mathbf{x})=m_{e}v_{the}^{2}(\mathbf{x})/2.$
Notice that, simply by balancing $eF_{0e}T_{0e}^{-1}\partial_{t}\varphi\sim\mathbf{v}_{E}\cdot\nabla F_{0e},$
we obtain 
\begin{equation}
\omega\sim\omega_{*e}=\frac{1}{2}k_{y}v_{the}\frac{\rho_{e}}{L_{n}},\label{eq:drift}
\end{equation}
 where $L_{n}^{-1}\sim F_{0e}^{-1}\nabla F_{0e}.$ Therefore, we now
include the electron drift frequency. Since in our fundamental ordering
we have $\omega\sim k_{\parallel}v_{the,}$ together with Eq. (\ref{eq:drift}),
this yields 
\begin{equation}
\frac{\rho_{e}}{L_{n}}\sim\frac{k_{\parallel}}{k_{\perp}}\equiv\epsilon.\label{eq:electr_ord}
\end{equation}
 The kinetic equation (\ref{eq:edk}), when linearized, corresponds
to that of Connor et al. in Ref \citep{0741-3335-51-1-015009}.

As in the homogeneous case \citep{zocco:102309}, we introduce a formal
mass ratio expansion for the electron gyrocentre distribution, so
that to zeroth order 
\begin{equation}
h_{e}=\left(-\frac{e\varphi}{T_{0e}}+\frac{\delta n_{e}}{n_{0e}}+\frac{v_{\parallel}u_{\parallel e}}{T_{0e}}m_{e}\right)F_{0e}+g_{e}+\mathcal{O}\left(\frac{m_{e}}{m_{i}}\right),\label{eq:split}
\end{equation}
here $u_{\parallel e}=n_{0e}^{-1}\int d^{3}\mathbf{v}v_{\parallel}h_{e}$,
and $\int d^{3}\mathbf{v}(1,v_{\parallel})g_{e}\equiv0.$ Using expression
(\ref{eq:split}) in Eq. (\ref{eq:edk}), and taking the zeroth moment
we obtain the electron continuity equation 
\begin{equation}
\frac{d}{dt}\frac{\delta n_{e}}{n_{0e}}+\hat{\mathbf{b}}\cdot\nabla u_{\parallel e}=-\frac{\mathbf{v}_{E}\cdot\nabla n_{0e}}{n_{0e}}\label{eq:elcontapp}
\end{equation}
 with $d/dt=\partial_{t}+\mathbf{v}_{E}\cdot\nabla.$ We recall that
every term kept in Eq. (\ref{eq:elcontapp}) is of order $\sim\epsilon/\sqrt{\beta_{e}}.$

After calculating the first moment of Eq. (\ref{eq:edk}) we have
the generalized Ohm's law 
\begin{equation}
\begin{split} & \frac{d}{dt}(A_{\parallel}-d_{e}^{2}\nabla_{\perp}^{2}A_{\parallel})=-c\frac{\partial\varphi}{\partial z}+\frac{T_{0e}c}{e}\hat{\mathbf{b}}\cdot\nabla\left[\frac{\delta n_{e}}{n_{0e}}+\frac{\delta T_{\parallel e}}{T_{0e}}\right]\\
 & -\frac{m_{e}c}{e}\frac{1}{n_{0e}}\int d^{3}\mathbf{v}v_{\parallel}\left(\frac{\partial h_{e}}{\partial t}\right)_{coll}+\frac{T_{0e}c}{e}\left(1+\eta_{e}\right)\frac{\tilde{\mathbf{b}}\cdot\nabla n_{0e}}{n_{0e}}\\
 & +\frac{m_{e}c}{e}\frac{1}{n_{0e}}\frac{d}{dt}n_{0e}u_{\parallel i},
\end{split}
\label{eq:grnohmslawapp}
\end{equation}
 with 
\begin{equation}
\eta_{e}=\frac{n_{oe}}{T_{0e}}\frac{\nabla T_{0e}}{\nabla n_{0e}},
\end{equation}

\begin{equation}
\frac{\delta T_{\parallel e}}{T_{0e}}\equiv\frac{1}{n_{0e}}\int d^{3}\mathbf{v}2\frac{v_{\parallel}^{2}}{v_{the}^{2}}g_{e},
\end{equation}
 and $\tilde{\mathbf{b}}\cdot\nabla=-B_{0}^{-1}\{A_{\parallel},\cdot\}.$
To derive Eq. (\ref{eq:grnohmslawapp}), we used the fact that 
\begin{equation}
u_{\parallel e}=\frac{e}{m_{e}c}d_{e}^{2}\nabla_{\perp}^{2}A_{\parallel}+u_{\parallel i},
\end{equation}
 and 
\begin{equation}
\nabla F_{0e}=F_{0e}\frac{\nabla n_{0e}}{n_{0e}}\left[1+\eta_{e}\left(\frac{m_{e}v^{2}}{2T_{0e}}-\frac{3}{2}\right)\right].
\end{equation}
 We can obtain an equation for $g_{e}$ after inserting Eqs. (\ref{eq:elcontapp})
and (\ref{eq:grnohmslawapp}) into (\ref{eq:edk}). The result is
\begin{equation}
\begin{split} & \frac{dg_{e}}{dt}+v_{\parallel}\left[\hat{\mathbf{b}}\cdot\nabla g_{e}-F_{0e}\hat{\mathbf{b}}\cdot\nabla\frac{\delta T_{\parallel e}}{T_{0e}}\right]-C[g_{e}]=\\
 & F_{0e}\left(1-2\frac{v_{\parallel}^{2}}{v_{the}^{2}}\right)\hat{\mathbf{b}}\cdot\nabla\left[\left(\frac{e}{m_{e}c}d_{e}^{2}\nabla_{\perp}^{2}A_{\parallel}+u_{\parallel i}\right)\right]+\\
 & -\eta_{e}F_{0e}\left(\frac{m_{e}v^{2}}{2T_{0e}}-\frac{3}{2}\right)\frac{\mathbf{v}_{E}\cdot\nabla n_{0e}}{n_{0e}}+\\
 & -\eta_{e}F_{0e}v_{\parallel}\left(\frac{m_{e}v^{2}}{2T_{0e}}-\frac{5}{2}\right)\frac{v_{the}\tilde{\mathbf{b}}\cdot\nabla n_{0e}}{n_{0e}},
\end{split}
\label{eq:krehminhom}
\end{equation}
 with the notation 
\begin{equation}
C[g_{e}]=\left(\frac{\partial h_{e}}{\partial t}\right)_{coll}-2\frac{v_{\parallel}F_{0e}}{v_{the}^{2}n_{oe}}\int d^{3}\mathbf{v}v_{\parallel}\left(\frac{\partial h_{e}}{\partial t}\right)_{coll}.
\end{equation}
 Equation (\ref{eq:krehminhom}), in the limit of a homogeneous background,
reduces to the result of Ref. \citep{zocco:102309}.

\subsection{Ion response and closure}

Equations (\ref{eq:elcontapp})-(\ref{eq:grnohmslawapp})-(\ref{eq:krehminhom})
{[}with (\ref{eq:krehminhom}) replaced by (\ref{eq:tempeq}) in the
collisional limit{]} are a set of three equations for $A_{\parallel},$
$\varphi,$ $g_{e}$ (which gives $\delta T_{\parallel e}/T_{0e}$),
and $\delta n_{e}/n_{0e}.$ To close the system we need an explicit
expression for the electron density perturbation $\delta n_{e}/n_{0e}.$
This can be readily calculated from the ion gyrokinetic equation.
Quasineutrality requires that $\delta n_{e}=\delta n_{i}.$ Under
the same orderings that produced Eq. (\ref{eq:krehminhom}), we have$^{1}$
\footnotetext[1]{These orderings are discussed at length in Ref.
\citep{zocco:102309}. What is evident is that streaming terms are
downgraded compared to those in the electron equation, because $v_{\parallel}\sim v_{thi}\sim(m_{e}/m_{i})^{1/2}v_{the}.$
Ions do not experience collisions, because they are too heavy to be
scattered by electrons {[}ion self-collisions will be eventually considered
in Eq. (A22){]}.} 
\begin{equation}
\frac{\partial h_{i}}{\partial t}+\left\langle \mathbf{v}_{E}\right\rangle \cdot\nabla h_{i}=\frac{ZeF_{0i}}{T_{0i}}\frac{\partial\left\langle \varphi\right\rangle }{\partial t}\,-\frac{c}{B_{0}}\mathbf{e}_{z}\cdot\nabla\left\langle \varphi\right\rangle \times\nabla F_{0i},\,\label{eq:iongk}
\end{equation}
 where the bracket has the usual meaning of a gyroaverage.

Let us introduce the function $g_{i}$ such that 
\begin{equation}
h_{i}=\frac{Ze\left\langle \varphi\right\rangle _{\mathbf{R}_{i}}}{T_{0i}}F_{0i}+g_{i}.
\end{equation}
 Thus, we obtain an inhomogeneous equation for $g_{i}$ 
\begin{equation}
\frac{dg_{i}}{dt}=-\frac{\left\langle \mathbf{v}_{E}\right\rangle \cdot\nabla n_{0i}}{n_{0i}}\left\{ 1+\eta_{i}\left(\hat{v}^{2}-\frac{3}{2}\right)\right\} F_{0i}.
\end{equation}
 This is a $5D$ equation, however, we can integrate over $\int dv_{\parallel}$
to obtain 
\begin{equation}
\begin{split} & \frac{\partial\hat{g}_{i}}{\partial t}+\frac{c}{B_{0}}\left\{ \left\langle \varphi\right\rangle _{\mathbf{R}_{i}},\,\hat{g}_{i}\right\} =\\
 & +i\sum_{\mathbf{k}^{\prime}}\omega_{*i}(k_{y}^{\prime})J_{0}(\mathbf{k}_{\perp}^{\prime}\rho_{i}\hat{v}_{\perp})\frac{Ze\varphi_{\mathbf{k}^{\prime}}}{T_{0i}}\left\{ 1-\eta_{i}+\eta_{i}\hat{v}_{\perp}^{2}\right\} \frac{n_{0i}e^{-\hat{v}_{\perp}^{2}}}{v_{thi}^{2}\pi}.
\end{split}
\label{eq:ionkin}
\end{equation}
where $\hat{g}_{i}=\int d\hat{v}_{\parallel}g_{i},$ and $\hat{v}=v/v_{thi}.$
When linearized, the ion response is the same as calculated in Ref.
\citep{antonsen-coppi,pegoraro:478,kadomtsev-pogutse,0741-3335-54-3-035003};
thus, in this case, we have 
\begin{equation}
\frac{\delta n_{i}}{n_{0i}}=\int_{-\infty}^{+\infty}dpe^{ipx}F(p\rho_{i})\frac{Ze\varphi}{T_{0i}}\equiv\hat{F}\frac{Ze\varphi_{\mathbf{k}}}{T_{0i}},\label{eq:GKPL}
\end{equation}
 with 
\begin{equation}
F(p\rho_{i})=-(1-\Gamma_{0})-\frac{\omega_{*i}}{\omega}\left[\Gamma_{0}+\frac{\eta_{i}}{2}p^{2}\rho_{i}^{2}(\Gamma_{0}-\Gamma_{1})\right],
\end{equation}
 $\omega_{*i}=-1/2k_{y}v_{thi}\rho_{i}/L_{n}\equiv-\tau\omega_{*e}<0,$
$\tau=T_{0i}/T_{0e},$ $\Gamma_{n}=\exp[-p^{2}\rho_{i}^{2}/2]I_{n}(p^{2}\rho_{i}^{2}/2),$
where $I_{n}$ is the modified Bessel function \citep{abram}. The
``hat'' on $F(p\rho_{i})$ is a short-hand notation for the inverse
transform. Notice that we are using $n_{0}(x)\approx-n_{0i}x/L_{n},$
as a local approximation for the equilibrium density profile.

A useful way to describe ion kinetics is the following. Let us introduce
the representation 
\begin{equation}
\hat{g}_{i}\left(\mathbf{R}_{i},\mathbf{v}_{\perp},t\right)=\sum_{\mathbf{k}}\sum_{n=0}^{\infty}g_{\mathbf{k}}^{n}L_{n}\left(\hat{v}_{\perp}^{2}\right)F_{0i}\left(\hat{v}_{\perp}^{2}\right)e^{i\mathbf{R}_{i}\cdot\mathbf{k}},\label{eq:ionrepre}
\end{equation}
 where $L_{n}$ are the Laguerre polynomials defined through the Rodriguez
formula. The velocity space representation is constructed on top of
the usual Fourier space representation, here written symbolically
as a summation. The coefficients $g_{\mathbf{k}}^{n}$ are thus defined
as 
\begin{equation}
g_{\mathbf{k}}^{n}=\frac{\pi}{n_{0i}}\int d\mathbf{R}_{i}\int_{0}^{\infty}dv_{\perp}v_{\perp}L_{n}\left(\hat{v}_{\perp}^{2}\right)\hat{g}_{i}\left(\mathbf{R}_{i},\mathbf{v}_{\perp},t\right)e^{-i\mathbf{R}_{i}\cdot\mathbf{k}}.
\end{equation}
 Using Eq. (\ref{eq:ionrepre}) to replace for $\hat{g}_{i}$ in Eq.
(\ref{eq:ionkin}), and we operating with $\pi/n_{0i}\int dv_{\perp}v_{\perp}L_{m}\left(\hat{v}_{\perp}^{2}\right),$
all the velocity space integrals can be carried out to obtain 
\begin{equation}
\begin{split} & \frac{\partial}{\partial t}g_{\mathbf{k}^{\prime}}^{n}-\frac{c}{B_{0}}\sum_{\mathbf{k}}^{\infty}\sum_{m=0}^{\infty}(-1)^{m+n}\mathbf{z}\cdot\mathbf{k}\times\mathbf{k}^{\prime}\\
 & \times\frac{1}{2}e^{-\frac{k_{\perp}^{2}\rho_{i}^{2}}{4}}L_{n}^{m-n}\left(\frac{1}{4}k_{\perp}^{2}\rho_{i}^{2}\right)\varphi_{\mathbf{k}}\times\\
 & L_{m}^{n-m}\left(\frac{1}{4}k_{\perp}^{2}\rho_{i}^{2}\right)g_{\mathbf{k}^{\prime}-\mathbf{k}}^{n}=\\
 & \omega_{*i}\frac{1}{2}e^{-\frac{k_{\perp}^{\prime2}\rho_{i}^{2}}{4}}\left[L_{n}\left(\frac{1}{4}k_{\perp}^{\prime2}\rho_{i}^{2}\right)L_{0}\left(\frac{1}{4}k_{\perp}^{\prime2}\rho_{i}^{2}\right)\,\right.\\
 & \left.-\eta_{i}\, L_{n}\left(\frac{1}{4}k_{\perp}^{\prime2}\rho_{i}^{2}\right)L_{1}\left(\frac{1}{4}k_{\perp}^{\prime2}\rho_{i}^{2}\right)\right]\frac{Ze\varphi_{\mathbf{k}^{\prime}}}{T_{0i}}.
\end{split}
\end{equation}

We expect nonlinear phase mixing to play a role now, and to create
structures in perpendicular velocity space for the ion distribution
function. Hence ion-ion collisions will eventually become important
for sufficiently large gradients $v_{thi}^{2}\partial_{v_{\perp}}^{2}\sim\omega/\nu_{ii}\gg1$
and a simple model collisional operator will need to be considered.
This aspect is not crucial for the scope of this paper.

We can now summarize the new set of equations 
\begin{equation}
\begin{split} & \frac{d}{dt}(A_{\parallel}-d_{e}^{2}\nabla_{\perp}^{2}A_{\parallel})=-c\frac{\partial\varphi}{\partial z}\\
 & \frac{T_{0e}c}{e}\hat{\mathbf{b}}\cdot\nabla\left[\frac{Z}{\tau}(\hat{\Gamma}_{0}-1)\frac{e\varphi_{\mathbf{k}}}{T_{0e}}+g_{i}^{(0)}+\frac{\delta T_{\parallel e}}{T_{0e}}\right]\\
 & \eta\nabla_{\perp}^{2}A_{\parallel}+\frac{T_{0e}c}{e}\left(1+\eta_{e}\right)\frac{\tilde{\mathbf{b}}\cdot\nabla n_{0e}}{n_{0e}},
\end{split}
\label{eq:sum1}
\end{equation}

\begin{equation}
\begin{split} & \frac{d}{dt}\left[\frac{Z}{\tau}(\hat{\Gamma}_{0}-1)\frac{e\varphi_{\mathbf{k}}}{T_{0e}}+g_{i}^{(0)}\right]+\hat{\mathbf{b}}\cdot\nabla\frac{e}{m_{e}c}d_{e}^{2}\nabla_{\perp}^{2}A_{\parallel}=\\
 & -\frac{\mathbf{v}_{E}\cdot\nabla n_{0e}}{n_{0e}},
\end{split}
\label{eq:sum2}
\end{equation}

\begin{equation}
\begin{split} & \frac{dg_{e}}{dt}+v_{\parallel}\left[\hat{\mathbf{b}}\cdot\nabla g_{e}-F_{0e}\hat{\mathbf{b}}\cdot\nabla\frac{\delta T_{\parallel e}}{T_{0e}}\right]-C[g_{e}]=\\
 & F_{0e}\left(1-2\frac{v_{\parallel}^{2}}{v_{the}^{2}}\right)\hat{\mathbf{b}}\cdot\nabla\left[\left(\frac{e}{m_{e}c}d_{e}^{2}\nabla_{\perp}^{2}A_{\parallel}+u_{\parallel i}\right)\right]+\\
 & -\eta_{e}F_{0e}\left(\frac{m_{e}v^{2}}{2T_{0e}}-\frac{3}{2}\right)\frac{\mathbf{v}_{E}\cdot\nabla n_{0e}}{n_{0e}}+\\
 & -\eta_{e}F_{0e}v_{\parallel}\left(\frac{m_{e}v^{2}}{2T_{0e}}-\frac{5}{2}\right)\frac{v_{the}\tilde{\mathbf{b}}\cdot\nabla n_{0e}}{n_{0e}},
\end{split}
\label{eq:sum3app}
\end{equation}
 with 
\begin{equation}
\frac{\delta T_{\parallel e}}{T_{0e}}\equiv\frac{1}{n_{0e}}\int d^{3}\mathbf{v}2\frac{v_{\parallel}^{2}}{v_{the}^{2}}g_{e},
\end{equation}
 and 
\begin{equation}
\begin{split} & g_{i}^{(0)}(\mathbf{r},t)=\frac{1}{n_{0i}}\int d^{3}\mathbf{v}\left\langle g_{i}(\mathbf{R}_{i},\mathbf{v}_{\perp},t)\right\rangle _{\mathbf{r}}\\
 & =\sum_{k}\sum_{n=0}^{\infty}\frac{\left(k_{\perp}\rho_{i}\right)^{n}}{2^{2n}n!}e^{-\frac{1}{4}k_{\perp}^{2}\rho_{i}^{2}}g_{\mathbf{k}}^{n}e^{i\mathbf{k}\cdot\mathbf{r}}.
\end{split}
\end{equation}
 We calculated explicitly the collision term in Ohm's law (\ref{eq:sum1}),
which gives the resistive contribution, with $\eta=\nu_{ei}d_{e}^{2}$
the Spitzer resistivity. We used the fact that 
\begin{equation}
\int_{0}^{\infty}dxxe^{-x^{2}}L_{n}(x^{2})J_{0}(xy)=\frac{2^{-2n}}{2n!}y^{2n}e^{-\frac{1}{4}y^{2}}.
\end{equation}
 This result can be proved by using the representation of Bessel functions
in terms of Laguerre polynomials 
\begin{equation}
J_{0}(k_{\perp}\rho_{i}\hat{v}_{\perp})=e^{-\frac{1}{4}k_{\perp}^{2}\rho_{i}^{2}}\sum_{n=0}^{\infty}\frac{\left(k_{\perp}^{2}\rho_{i}^{2}/4\right)^{n}}{n!}L_{n}(\hat{v}_{\perp}^{2}),
\end{equation}
 and the orthogonality of these polynomials. We also used the fact
that \citep{erdelyi2} 
\begin{equation}
\begin{split} & \int_{0}^{\infty}dxxe^{-x^{2}}L_{n}(x^{2})L_{m}(x^{2})J_{0}(xy)=\\
 & \frac{(-1)^{m+n}}{2}e^{-\frac{1}{4}y^{2}}L_{n}^{m-n}(\frac{y^{2}}{4})L_{m}^{n-m}(\frac{y^{2}}{4}).
\end{split}
\label{eq:integral}
\end{equation}
 The integral (\ref{eq:integral}) is a simplified version of the
integral \citep{gradshteyn-ryzhik} 
\begin{equation}
\begin{split} & I_{m,n}=\int_{0}^{\infty}dxx^{\nu+1}e^{-\alpha x^{2}}L_{m}^{\nu-\sigma}(\alpha x^{2})L_{n}^{\sigma}(\alpha x^{2})J_{\nu}(xy)=\\
 & \frac{(-1)^{m+n}}{2}e^{-\frac{1}{4}y^{2}}L_{n}^{m-n-\sigma}(\frac{y^{2}}{4})L_{m}^{n-m+\sigma-\nu}(\frac{y^{2}}{4}),
\end{split}
\end{equation}
 which is the same as given in Ref. \citep{erdelyi2}. In Ref. \citep{gradshteyn-ryzhik}
we find the following conditions: $n\neq0,$ $\sigma\neq0,$ $\alpha\neq1.$
However, if we set $n=0,$ $\sigma=0,$ $\alpha=1,$ \emph{and} $\nu=0,$
we calculate analytically term by term for each $m$: 
\begin{equation}
\begin{split}m=0 & \,\,\,\, I_{m,0}=\frac{1}{2}e^{-\frac{1}{4}y^{2}}\\
m=1 & \,\,\, I_{m,0}=\frac{1}{32}e^{-\frac{1}{4}y^{2}}(y^{2}-4)^{2}\\
m=1 & \,\,\, I_{m,0}=\frac{1}{2048}e^{-\frac{1}{4}y^{2}}(y^{4}-16y^{2}+32)^{2}...,\mbox{et\,\ cetera.}
\end{split}
\end{equation}
 Notice that, by construction of the ordering, $g_{i}$ is only a
function of $v_{\perp},$ thus the ions are not carrying any parallel
current. \vspace{3cm}

\vspace{10cm}

\bibliographystyle{IEEEtran} \bibliographystyle{iopart-num}
\bibliography{MT_final_23_10_14}

\end{document}